\documentclass[twocolumn,onecolappendix]{aastex631}
\usepackage{amsmath}
\shorttitle{eDisk XIV. Modeling of a Protostellar Disk}
\shortauthors{Takakuwa et al.}
\graphicspath{{./}{figures/}}

\begin{document}
\title{Early Planet Formation in Embedded Disks (eDisk) XIV:\\
Flared Dust Distribution and Viscous Accretion Heating
of the Disk around R~CrA~IRS~7B-a}

\correspondingauthor{Shigehisa Takakuwa}
\email{takakuwa@sci.kagoshima-u.ac.jp}

\author[0000-0003-0845-128X]{Shigehisa Takakuwa}
\affiliation{Department of Physics and Astronomy, Graduate School of Science and Engineering, Kagoshima University, 1-21-35 Korimoto, Kagoshima, Kagoshima 890-0065, Japan}
\affiliation{Academia Sinica Institute of Astronomy \& Astrophysics,
11F of Astronomy-Mathematics Building, AS/NTU, No.1, Sec. 4, Roosevelt Rd, Taipei 10617, Taiwan, R.O.C.}

\author[0000-0003-1549-6435]{Kazuya Saigo}
\affiliation{Department of Physics and Astronomy, Graduate School of Science and Engineering, Kagoshima University, 1-21-35 Korimoto, Kagoshima, Kagoshima 890-0065, Japan}

\author[0000-0002-2902-4239]{Miyu Kido}
\affiliation{Department of Physics and Astronomy, Graduate School of Science and Engineering, Kagoshima University, 1-21-35 Korimoto, Kagoshima, Kagoshima 890-0065, Japan}

\author[0000-0003-0998-5064]{Nagayoshi Ohashi}
\affiliation{Academia Sinica Institute of Astronomy \& Astrophysics,
11F of Astronomy-Mathematics Building, AS/NTU, No.1, Sec. 4, Roosevelt Rd, Taipei 10617, Taiwan, R.O.C.}

\author[0000-0002-6195-0152]{John J. Tobin}
\affil{National Radio Astronomy Observatory, 520 Edgemont Rd., Charlottesville, VA 22903 USA} 

\author[0000-0001-9133-8047]{Jes K. J{\o}rgensen}
\affil{Niels Bohr Institute, University of Copenhagen, {\O}ster Voldgade 5--7, DK~1350 Copenhagen K., Denmark}

\author[0000-0003-3283-6884]{Yuri Aikawa}
\affiliation{Department of Astronomy, Graduate School of Science,
The University of Tokyo, 7-3-1 Hongo, Bunkyo-ku, Tokyo 113-0033, Japan}

\author[0000-0002-8238-7709]{Yusuke Aso}
\affiliation{Korea Astronomy and Space Science Institute, 776 Daedeok-daero, Yuseong-gu, Daejeon 34055, Republic of Korea}



\author[0000-0001-5782-915X]{Sacha Gavino}
\affiliation{Niels Bohr Institute, University of Copenhagen, {\O}ster Voldgade 5--7, DK~1350 Copenhagen K., Denmark}


\author[0000-0002-9143-1433]{Ilseung Han}
\affiliation{Division of Astronomy and Space Science, University of Science and Technology, 217 Gajeong-ro, Yuseong-gu, Daejeon 34113, Republic of Korea}
\affiliation{Korea Astronomy and Space Science Institute, 776 Daedeok-daero, Yuseong-gu, Daejeon 34055, Republic of Korea}



\author[0000-0003-2777-5861]{Patrick M. Koch}
\affiliation{Academia Sinica Institute of Astronomy \& Astrophysics,
11F of Astronomy-Mathematics Building, AS/NTU, No.1, Sec. 4, Roosevelt Rd, Taipei 10617, Taiwan, R.O.C.}

\author[0000-0003-4022-4132]{Woojin Kwon}
\affiliation{Department of Earth Science Education, Seoul National University, 1 Gwanak-ro, Gwanak-gu, Seoul 08826, Republic of Korea}
\affiliation{SNU Astronomy Research Center, Seoul National University, 1 Gwanak-ro, Gwanak-gu, Seoul 08826, Republic of Korea}


\author[0000-0002-3179-6334]{Chang Won Lee}
\affiliation{Division of Astronomy and Space Science, University of Science and Technology, 217 Gajeong-ro, Yuseong-gu, Daejeon 34113, Republic of Korea}
\affiliation{Korea Astronomy and Space Science Institute, 776 Daedeok-daero, Yuseong-gu, Daejeon 34055, Republic of Korea}

\author[0000-0003-3119-2087]{Jeong-Eun Lee}
\affiliation{Department of Physics and Astronomy, Seoul National University, 1 Gwanak-ro, Gwanak-gu, Seoul 08826, Korea}

\author[0000-0002-7402-6487]{Zhi-Yun Li}
\affiliation{University of Virginia, 530 McCormick Rd., Charlottesville, Virginia 22904, USA}

\author[0000-0001-7233-4171]{Zhe-Yu Daniel Lin}
\affiliation{University of Virginia, 530 McCormick Rd., Charlottesville, Virginia 22904, USA}

\author[0000-0002-4540-6587]{Leslie W.  Looney}
\affiliation{Department of Astronomy, University of Illinois, 1002 West Green St, Urbana, IL 61801, USA}

\author[0000-0002-7002-939X]{Shoji Mori}
\affiliation{Astronomical Institute, Graduate School of Science, Tohoku University, Sendai 980-8578, Japan}





\author[0000-0003-4361-5577]{Jinshi Sai (Insa Choi)}
\affiliation{Academia Sinica Institute of Astronomy \& Astrophysics,
11F of Astronomy-Mathematics Building, AS/NTU, No.1, Sec. 4, Roosevelt Rd, Taipei 10617, Taiwan, R.O.C.}


\author[0000-0002-0549-544X]{Rajeeb Sharma}
\affiliation{Niels Bohr Institute, University of Copenhagen, {\O}ster Voldgade 5--7, DK~1350 Copenhagen K., Denmark}

\author[0000-0002-9209-8708]{Patrick Sheehan}
\affiliation{National Radio Astronomy Observatory, 520 Edgemont Rd., Charlottesville, VA 22903 USA}


\author[0000-0001-8105-8113]{Kengo Tomida}
\affiliation{Astronomical Institute, Graduate School of Science, Tohoku University, Sendai 980-8578, Japan}


\author[0000-0001-5058-695X]{Jonathan P. Williams}
\affiliation{Institute for Astronomy, University of Hawai‘i at Mānoa, 2680 Woodlawn Dr., Honolulu, HI 96822, USA}

\author[0000-0003-4099-6941]{Yoshihide Yamato}
\affiliation{Department of Astronomy, Graduate School of Science, The University of Tokyo, 7-3-1 Hongo, Bunkyo-ku, Tokyo 113-0033, Japan}

\author[0000-0003-1412-893X]{Hsi-Wei Yen}
\affiliation{Academia Sinica Institute of Astronomy \& Astrophysics,
11F of Astronomy-Mathematics Building, AS/NTU, No.1, Sec. 4, Roosevelt Rd, Taipei 10617, Taiwan, R.O.C.}




\begin{abstract}

We performed radiative transfer calculations and observing simulations
to reproduce the
1.3-mm dust-continuum and C$^{18}$O (2--1) images in
the Class I protostar R CrA IRS7B-a, observed with the ALMA Large Program
``Early Planet Formation in Embedded Disks (eDisk)". We found that
the dust disk model passively heated by the central protostar cannot
reproduce the observed peak brightness temperature of the 1.3-mm continuum emission ($\sim$195~K), regardless of the assumptions about the dust opacity.
Our calculation suggests that viscous accretion heating in the disk is 
required to reproduce
the observed high brightness temperature. The observed intensity profile
of the 1.3-mm dust-continuum emission along the disk minor axis is skewed
toward the disk far side.
Our modeling reveals that such an asymmetric
intensity distribution requires flaring of the dust along the disk's
vertical direction with the scale-height following $h/r \sim r^{0.3}$ as function of radius.
These results are in sharp contrast
to those of Class II disks,
which show geometrically flat dust distributions and lower dust temperatures.
From our modeling of the
C$^{18}$O (2--1) emission, the outermost radius of the gas disk
is estimated to be $\sim$80~au, larger than that of the dust disk ($\sim$62~au),
to reproduce the observed distribution of the C$^{18}$O (2--1) emission in IRS~7B-a.
Our modeling unveils a hot and thick dust disk plus a larger gas disk
around one of the eDisk targets, which could be
applicable to other protostellar sources in contrast to more evolved sources.
\end{abstract}



\section{Introduction} \label{sec:intro}

\begin{figure*}[ht!]
\begin{center}
\includegraphics[width=180mm, angle=0]
{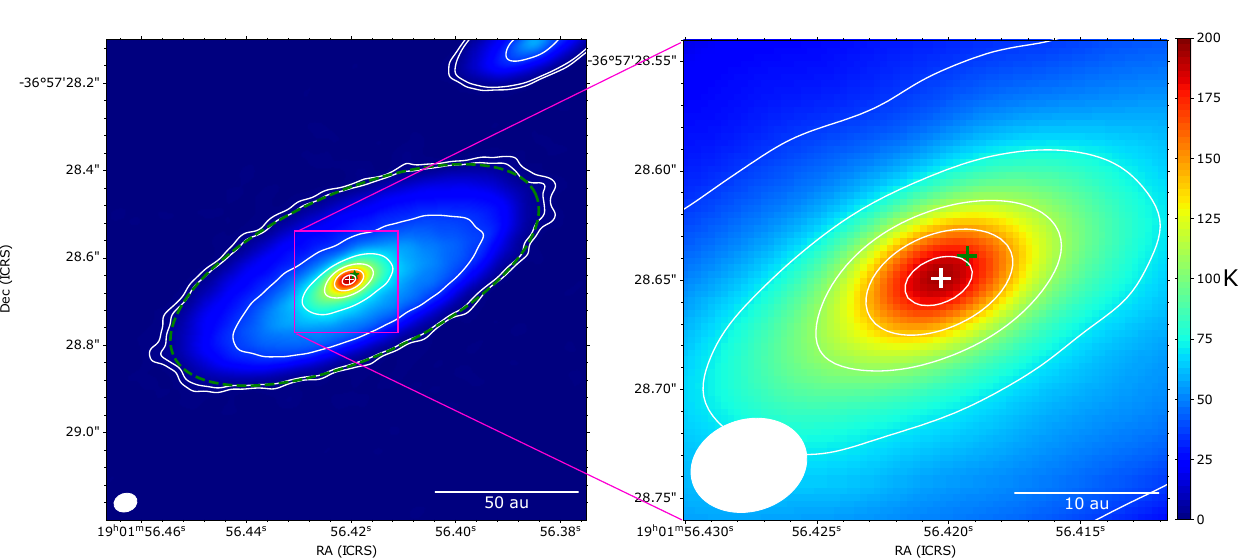}
\caption{Overall ($left$) and zoom-in images ($right$) of the observed 1.3-mm dust continuum emission in IRS~7B-a.
Solid contours show 5$\sigma$, 9$\sigma$, 100$\sigma$,
200$\sigma$, 300$\sigma$, 400$\sigma$, and 500$\sigma$ (1$\sigma$ = 0.033 mJy beam$^{-1}$ = 0.36 K).
The beam size is 0$\farcs$054 $\times$ 0$\farcs$042 (P.A. = -72.6$\degr$).
A white cross denotes the peak position of
the 1.3-mm dust continuum emission,
whose coordinates are (19$^h$01$^m$56$\fs$420, -36$\degr$57$\arcmin$28$\farcs$65).
This position is adopted
as the origin of the intensity profile and
the position-velocity diagrams.
For reference, a green cross is an approximate geometrical center of the ellipse (dashed green line), which delineates the contour with the intensity of 0.3 mJy beam$^{-1}$. 
\label{fig:dustobs}}
\end{center}
\end{figure*}

The ALMA Large Program ``Early Planet Formation in Embedded Disk (eDisk)''
has newly observed eleven Class 0 and six Class I protostars
in 1.3-mm dust continuum and selected molecular lines,
including C$^{18}$O (2--1), around 230~GHz at
a spatial resolution of $\sim$7~au (plus one Class 0 and I sources
from archival data).
The datasets enable us to investigate the ongoing disk and planet formation
occurring during the protostellar stages systematically \citep{2023Ohashi}.
As reported in the present series of the eDisk
first-look papers, the initial results have demonstrated intriguing features of
the protostellar disks, envelopes, and outflows. 
For example, eDisk has unveiled that Keplerian rotating disks
are often found even in the Class 0 stage
\citep{2023Hoff,2023Kido,2023Sharma,2023Aso,2023Sai,2023Thieme}.
While concentric ring and
gap features, often seen in Class II disks \citep[e.g.,][]{2018Andrews},
are only seen toward the most evolved eDisk sources of Oph IRS 63
\citep{2023Flores} and L1489 IRS \citep{2023Yamato},
``bump" or ``shoulder" features are seen in Class 0 sources of
Ced 110 IRS 4 \citep{2023Sai} and CB 68 \citep{2023Kido}.
In the envelopes, flow-like molecular gas accreting toward the central
protostellar disks, accretion streamers, have been identified
\citep{2023Kido,2023Aso,2023Flores,2023Hanedisk}. An extended circumbinary envelope
around the compact binary disks is also identified \citep{2023Narayanan}.

To put these observational results into a quantitative astrophysical context, we need to compare the observations to detailed synthetic observations based on radiative transfer modeling of protostellar disks and envelopes
\citep[see][]{2020Baek} properly processed with simulations of the 
ALMA imaging process.
In this paper, we report our first attempt to reproduce the observed
1.3-mm continuum and C$^{18}$O (2--1) emission in one of the eDisk targets; R~CrA~IRS~7B \citep{2007Groppi,2011Peterson,2012Lindberg,2014LindbergHerschel,2014LindbergIRS7B}.
R~CrA~IRS~7B is a Class I protostar located in Corona Australis, with a bolometric luminosity
of $L_{\rm bol}$ = 5.1 $L_{\odot}$ and a bolometric temperature of $T_{\rm bol}$ = 88 K,
which are reestimated using the archival photometric data by the eDisk team
\citep[see][for details]{2023Ohashi}.
Note that both in terms of its bolometric luminosity and temperature
R~CrA~IRS~7B is typical among the eDisk targets.
The distance to the source is estimated
to be 152 pc based on the GAIA data \citep{2020Galli}.
Previous SMA and ALMA observations of R~CrA~IRS~7B demonstrated the presence of compact 1.3-mm and 0.8-mm continuum emission \citep{2012Lindberg,2014LindbergIRS7B}.
Through ALMA observations of the C$^{17}$O (3--2) line, a velocity gradient along the major axis of the dust
emission was identified, which is interpreted as the signature of Keplerian rotation \citep{2014LindbergIRS7B}.
The angular resolutions of these previous observations are, however, not sufficient
to investigate the detailed internal structure of the protostellar disk.
The eDisk observations for the first time revealed that R~CrA~IRS~7B
consists of two sources;
one to the southeast (R~CrA~IRS~7B-a) and
the other to the northwest (R~CrA~IRS~7B-b), each surrounded by 
a separate dust disk \citep{2023Ohashi}.
In the present paper, we focus on the
southeastern source, R~CrA~IRS~7B-a, which
is referred to as IRS~7B-a hereafter.

Figure \ref{fig:dustobs} shows the entire and zoom-in views
of the observed 1.3-mm dust-continuum emission in IRS~7B-a
\citep{2023Ohashi}.
The peak brightness temperature of the 1.3-mm
dust-continuum emission is as high as $\sim$195~K. 
It is clear that
along the minor axis, the emission gradient toward
the northeast is steeper than toward the southwest, suggesting the presence of asymmetry
(Figure \ref{fig:dustobs} right).
\citet{2023Ohashi} also found that
the C$^{18}$O (2--1) emission shows a signature of the Keplerian rotation
in the disk with an inferred central protostellar mass of 2.1--3.2~$M_{\odot}$,
while the emission becomes weaker
at the disk center against the bright dust continuum emission.

Similar characteristics are also seen in several other
eDisk sources, and thus it is important to reproduce these features
with radiative transfer modeling and to study the physical causes
of such characteristics. Our approach is, however, not to construct elaborated
theoretical models of the protostellar system or perform quantitative fitting of the model to the observed images. Instead, we aim to construct a simple, but sufficient, physical model that can reproduce these observed characteristic features as quantitatively as possible. Such an
approach enables us to discuss the primary causes of the observed features and the important insights behind them. 

The structure of the present paper is as follows.
In Section \ref{sec:model} physical models of the 
disk and envelope are described, followed by a
detailed explanation of the radiative transfer calculations
and the model parameter search.
In Section \ref{sec:dustmodel} the model images of the 1.3-mm dust continuum emission with a range of
model parameters are compared
to the real observed 1.3-mm image. Physical causes of the
high observed brightness temperature of the
1.3-mm dust continuum emission and the asymmetric intensity profile
along the disk minor axis are discussed separately.
In Section \ref{sec:c18o} the model C$^{18}$O velocity channel maps
and P-V diagrams are compared to those of the real data.
In Section \ref{sec:discussion} we discuss implications of our modeling results
in the context of the evolution of disks into planet formation.
Section \ref{sec:sum} gives a concise summary of the present work.

\section{Model} \label{sec:model}
\begin{figure*}[ht!]
\centering
\includegraphics[width=130mm, angle=0]{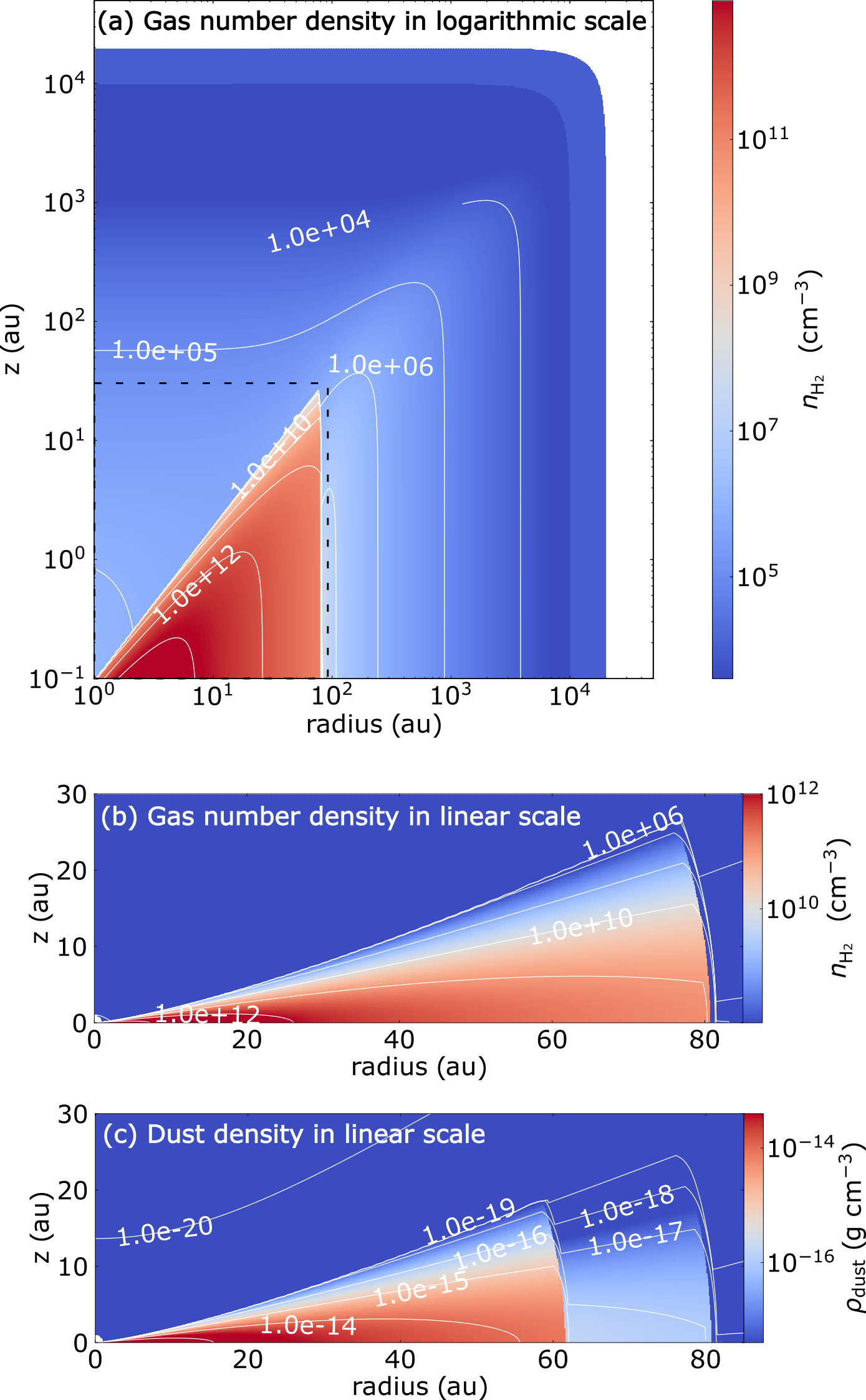}
\caption{a) Distribution of the gas number density ($\equiv n_{\rm H_2}$) of the fiducial model in the log $R$ - log $z$ plane.
The high-density region represents the disk, and the region outside
the disk represents the protostellar envelope. In the outermost part of the envelope,
a static, uniform cocoon with $n_{\rm H_2}$ = 10$^4$ cm$^{-3}$ is present.
Contour levels start from 10$^4$ cm$^{-3}$ in steps of a factor of 10.
A black dashed rectangle delineates
the zoom-in region shown in panels b and c.
b) Zoom-in view of the $n_{\rm H_2}$ distribution in the disk region in the $R - z$ plane.
Contour levels start from 10$^6$ cm$^{-3}$ in steps of a factor of 10.
c) Same as panel b) but for the distribution of the dust density, $\rho_{\rm dust}$.
Note the shorter outermost radius of the dust disk.
Contour levels start from 10$^{-20}$ g cm$^{-3}$ in steps of a factor of 10. 
\label{fig:modelconf}}
\end{figure*}

We introduce protostellar disk + envelope models which
should be able to capture the main observed features
with the minimum model complexity.
The gas and dust distributions are assumed to be azimuthally symmetric
as well as mirror-symmetric with respect to the midplane.
The spatial grids in a spherical polar coordinate system are
($r$, $\theta$, $\phi$) = (512, 512, 1).
In the radial direction, an equally spaced grid on a logarithmic scale is
adopted. 
In the polar direction a linearly equally-spaced grid is adopted.
In our preliminary parameter search, a coarser grid of
(256, 256, 1) is also adopted for efficient calculations.
The radius $r$ and elevation angle $\theta$ range 
$r$ = 1.0 -- 20000~au and $\theta$ = 0$\degr$ -- 90$\degr$, respectively.
The innermost radius of 1 au is fine enough for the spatial resolution
of the observed images ($\sim$7 au).
An even smaller inner radius
makes the gas temperature too high, and
the collisional coefficients of C$^{18}$O unavailable.
Within this defined space our model includes three distinct components;
a Keplerian disk, a rotating and infalling protostellar envelope
surrounding the disk, and
a static ``cocoon'' of molecular gas surrounding the envelope.
The cocoon is a spherical shell with a constant
gas density of \textcolor{black}{10$^4$ cm$^{-3}$ between $r$=10000~au and 20000~au that mimics the ambient
molecular cloud component.}
No gas motion except for an isotropic turbulence
of \textcolor{black}{0.2 km s$^{-1}$} is included in the cocoon.

\subsection{Keplerian Disk} \label{sec:disk}
The implementation of the Keplerian disk is similar to that by \citet{2015Kwon}
and \citet{2021Ichikawa}.
The disk properties are calculated with the cylindrical radius $R$,
\begin{equation}
R = r \sin\theta.
\end{equation}
The vertical scale height $h_0$ of dust and gas at $R = R_0$ is set to be
\begin{equation}
h_0 = \frac{c_{\rm s_0}}{\Omega_{\rm K_0}},
\end{equation}
where
\begin{equation}
c_{\rm s_0} = \sqrt{\frac{k_{\rm B} T_0}{m_{\rm p} \mu}},
\end{equation}
and
\begin{equation}
\Omega_{\rm K_0} = \sqrt{\frac{G M_{\star}}{R_0^3}},
\end{equation}
are the sound speed and the Keplerian angular velocity in the midplane, respectively,
at a reference radius $R_0=1$~au, and where $k_{\rm B}$ denotes the Boltzmann constant,
$T_0$ the midplane dust temperature at $R_0$, $m_{\rm p}$ the proton mass,
\textcolor{black}{$\mu$ (=2.33)} the mean molecular weight,
$G$ the gravitational constant, and $M_{\star}$ the stellar mass.
$M_{\star}$ is derived to be in the range of
$\sim$2.1 - 3.2 $M_{\odot}$ from the analysis of
the observed Position-Velocity diagram in the C$^{18}$O (2--1) emission
\citep{2023Ohashi}, using the SLAM package
\citep{yusuke_aso_2023_7783868}. In our model $M_{\star}$ = 2.9 $M_{\odot}$
is adopted, which provides a good match of the model velocity channel maps
with the observed velocity channel maps in the C$^{18}$O emission
(see section \ref{sec:c18o}).
\textcolor{black}{We set $T_0$ = 400 K, which}
is simply adopted to calculate the scale height. The true gas and dust temperature will be calculated later with a given density distribution.
The radial profile of the scale height ($\equiv h(R)$), or the disk flaring,
is expressed as;
\textcolor{black}{
\begin{equation}
\frac{h(R)}{R} = \frac{h_0}{R_0} \left( \frac{R}{R_0} \right)^{q},
\end{equation}
}
where $q$ denotes the disk flaring index.

The gas surface density profile $\Sigma(R)$ is assumed
to follow the power-law profile as
\begin{equation}
\Sigma (R) = \Sigma_{0} \left(\frac{R}{R_0}\right)^p,
\end{equation}
where $\Sigma_{0}$ is the surface density at $R$ = $R_0$, and $p$ is the power-law index.
$p$ is normally assumed to be in the range between -0.5 and -1, and here
\textcolor{black}{$p$ = -0.5} is adopted.
The disk is sharply truncated at the inner spherical
radius $r < r_{in}$ (= 1 au) and the outer radius $r > r_{gas}$.
$\Sigma_{0}$ is then derived from the disk mass ($\equiv M_d$) as
\textcolor{black}{
\begin{equation}
\Sigma_{0} = \frac{(2+p)R_0^p M_d}{2 \pi (r_{gas}^{2+p} - r_{in}^{2+p}) }.
\end{equation}
}
The gas mass density $\rho$ can be described as
\textcolor{black}{
\begin{equation}
\small
\rho(R, z) = \frac{\Sigma(R)}{h(R)\sqrt{2\pi}}
\exp\left({-\frac{1}{2}\left(\frac{z}{h(R)}\right)^2}\right).
\end{equation}
}
The velocity field in the disk is a simple Keplerian rotation, expressed as;
\begin{equation}
v_{\phi} = \sqrt{\frac{GM_{\star}}{r}}.
\end{equation}

The gas-to-dust mass ratio is assumed to be 100.
During our modeling effort to reproduce both the 1.3-mm dust-continuum
and C$^{18}$O (2--1) emission, we found, however, that the radius of the
dust disk is likely smaller than that of the gas disk
\textcolor{black}{(see Section \ref{sec:c18o})}.
Thus, we introduce an additional parameter, $r_{dust}$, the radius of
the dust disk. Between $r_{dust}$ and $r_{gas}$,
\textcolor{black}{the gas-to-dust mass ratio is arbitrarily
enhanced by a factor of 50}
to mimic a dust-free region of molecular gas (see section \ref{sec:param}).
The C$^{18}$O abundance is assumed to be the canonical
value of \textcolor{black}{1.76 $\times$ 10$^{-7}$} \citep{2004Crapsi}
and constant in the disk as well as in the envelope.
As shown below, the calculated dust temperature in the disk
is above the CO freeze-out temperature of 25 K in most
parts of the disk, and thus the constant CO abundance is a reasonable assumption.

\subsection{Protostellar Envelope} \label{sec:env}

In the radial range from $r_{gas}$ to \textcolor{black}{$r$ = 10000~au},
and above
\textcolor{black}{$z >$4 $\times$ $h$ at $r_{in} \leq r < r_{gas}$,}
the protostellar envelope of molecular gas and dust is filled.
The gas-to-dust mass ratio is assumed to be 100 throughout the envelope
as well as the cocoon.
The choice of the vertical boundary between the disk and envelope
at \textcolor{black}{4 $\times$ $h$} is somewhat arbitrary.
If the boundary is set to be 1 $\times$ $h$, the model disk
is too thin to show the effect of the disk flaring.
We also note that the inclusion of the envelope component is important,
even if the observed 1.3-mm dust-continuum and the C$^{18}$O (2--1) emission
appear to primarily trace the protostellar disk. As we will show below,
inclusion of viscous accretion heating is likely required to reproduce the
intense 1.3-mm dust-continuum emission originating from the disk.
The protostellar envelope should supply the accreting material to the disk.
Furthermore, the presence of the protostellar envelope acts as a ``blanket"
for the disk by slowing down the escape of photons from the disk.

The model of the rotating and infalling protostellar envelope is taken
from \citet{1976Ulrich,1994Hartmann}, and \citet{2004Mendoza}.
It is a ballistic solution to the two-body problem, with the envelope material following parabolic trajectories around the central protostar.
The velocity vector in the 3-dimensional space is expressed as:
\begin{equation}
v_r = -\left(\frac{GM_{\star}}{r}\right)^{\frac{1}{2}}\left(1+{\frac{\cos\theta}{\cos\theta_0}}\right)^{\frac{1}{2}},
\end{equation}
\begin{equation}
v_\theta = \left(\frac{GM_{\star}}{r}\right)^{\frac{1}{2}}\left(\frac{\cos\theta_0-\cos\theta}{\sin\theta}\right)\left(1+{\frac{\cos\theta}{\cos\theta_0}}\right)^{\frac{1}{2}},
\end{equation}
\begin{equation}
v_\phi = \left(\frac{GM_{\star}}{r}\right)^{\frac{1}{2}}\frac{\sin\theta_0}{\sin\theta}\left(1-{\frac{\cos\theta}{\cos\theta_0}}\right)^{\frac{1}{2}},
\end{equation}
where $\theta_0$ denotes the initial elevation angle of the infalling material.
Once the envelope material enters the disk boundary (see above), 
the material is assumed to be incorporated into the disk.
In the case of the isotropic infall in the envelope, the above infalling velocity yields
the expression of the envelope density as;
\begin{equation}
\begin{split}
\small
\rho(r,\theta) = \rho_0 \left(\frac{r}{r_{gas}} \right)^{-\frac{3}{2}} \left(1+{\frac{\cos\theta}{\cos\theta_0}}\right)^{-\frac{1}{2}} \\
\times \left(3\frac{r_{gas}}{r} \cos^2\theta_0 + 1 - \frac{r_{gas}}{r}\right)^{-1},
\end{split}
\end{equation}
where
\begin{equation}
\rho_0 = \frac{\dot{M}}{4\pi}\sqrt{\frac{1}{GM_{\star}r_{gas}^3}},
\end{equation}
and $\dot{M}$ indicates the mass infalling rate of the envelope.
Note that given $M_{\star}$ and $r_{gas}$, the density structure of the envelope is uniquely determined by $\dot{M}$.

The formulation proposed by \citet{1994Hartmann} is adopted
to express the flattening of the envelope as;
\begin{equation}
\rho_{\rm flat}(r,\theta) = \frac{\eta_f}{\cosh^{2}(\eta_f \cos\theta_0) \tanh \eta_f} \rho(r,\theta),
\end{equation}
where $\eta_f$ represents the degree of flattening of the envelope. $\eta_f$ = 1 denotes no flattening,
$i.e.,$ $\rho_{flat}(r,\theta) = \rho(r,\theta)$, while higher values of $\eta_f$
indicate more flattening. We adopt $\eta_f = 2$, following \citet{1998Momose}.

Figure \ref{fig:modelconf}a and b
show distributions of the volume gas density $n_{\rm H_2}$ (cm$^{-3}$)
in the envelope plus the cocoon and the disk, respectively, of our fiducial model.
Figure \ref{fig:modelconf}c shows the distribution of the dust density
in the disk. The difference of the radii between the molecular and dust
disks is presented.
The envelope mass of the fiducial model is 0.062 $M_{\odot}$, in contrast with
the disk gas mass of 0.41 $M_{\odot}$.
The cocoon mass is 1.9 $M_{\odot}$.

\subsection{Radiative Transfer Calculations} \label{sec:RADMC-3D}

\begin{figure*}[ht!]
\begin{center}
\includegraphics[width=130mm, angle=0]{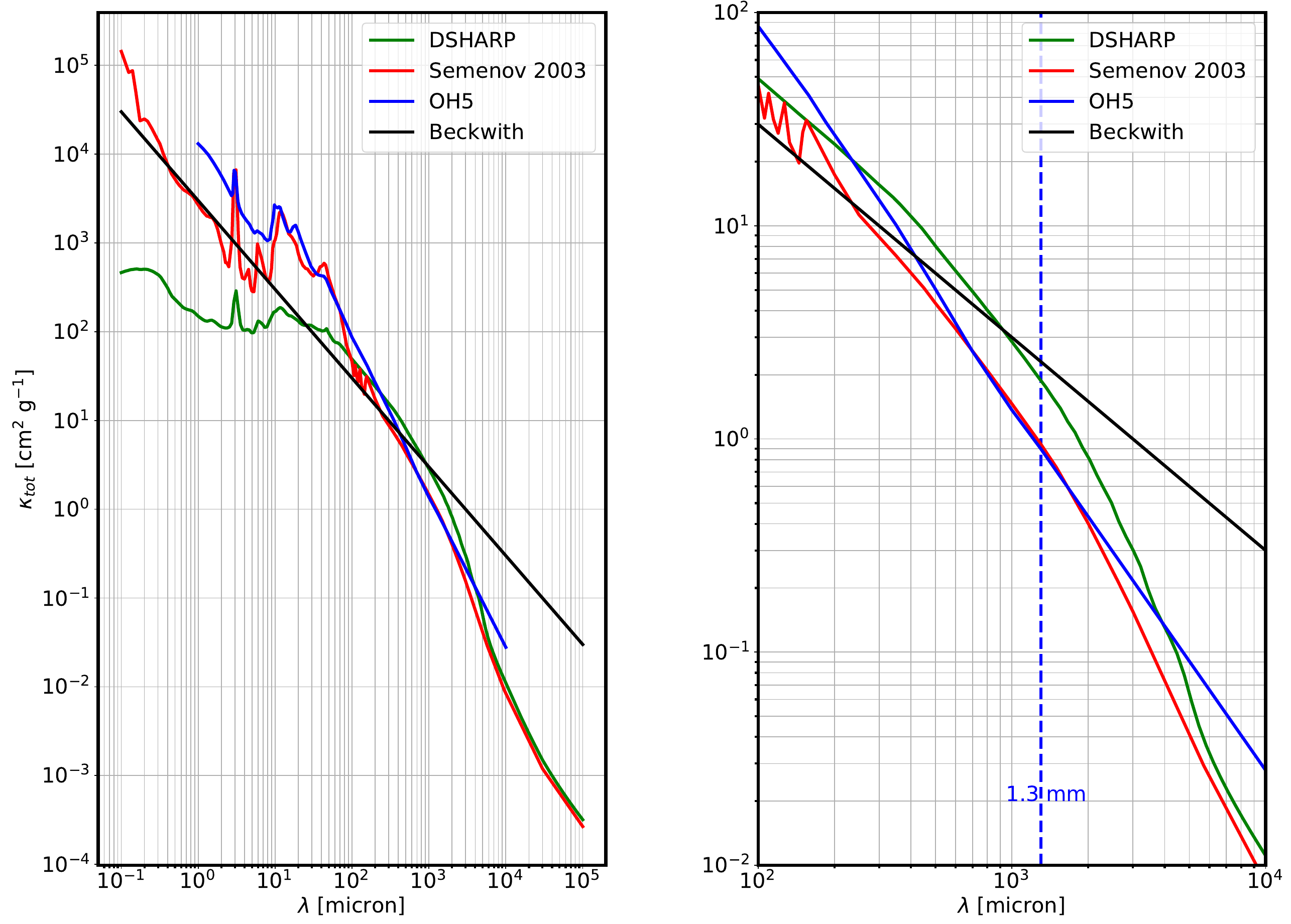}
\caption{(Left) Dust mass opacity (absorption and scattering) $\kappa_{tot}$ (cm$^2$ g$^{-1}$)
as a function of the wavelength of the four dust models; DSHARP, Semenov, OH5, and Beckwith. (Right) Zoom-in view of the left panel in the submillimeter
and millimeter wavelength.
\label{fig:kappa}}
\end{center}
\end{figure*}

\begin{figure*}[ht!]
\begin{center}
\includegraphics[width=150mm, angle=0]{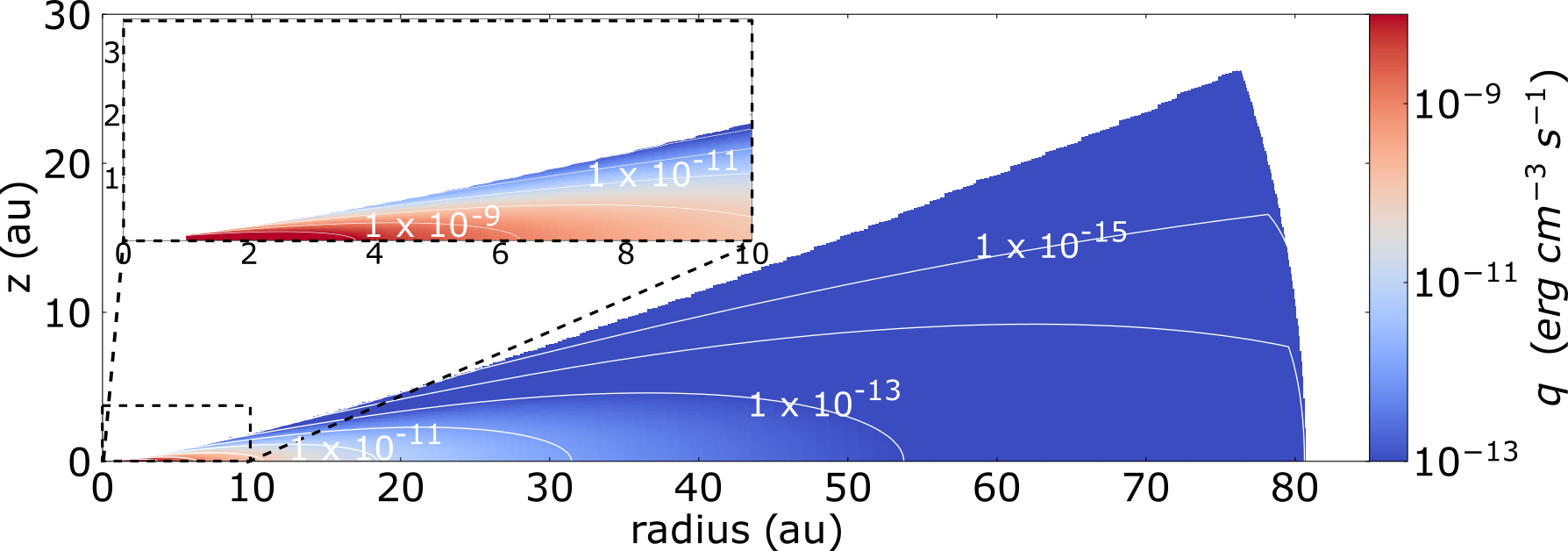}
\caption{Distribution of the internal heating rate $q_{heat}$ by the viscous accretion heating of the fiducial model, where
\textcolor{black}{$\dot{M}$ = 1.4$\times$10$^{-6}$ ($M_{\odot}~{\rm yr}^{-1}$)}.
\textcolor{black}{Contour levels start from 10$^{-15}$ erg cm$^{-3}$ s$^{-1}$ in steps of a factor of 10.}
\label{fig:visheat}}
\end{center}
\end{figure*}

\begin{figure*}[ht!]
\begin{center}
\includegraphics[width=100mm, angle=0]{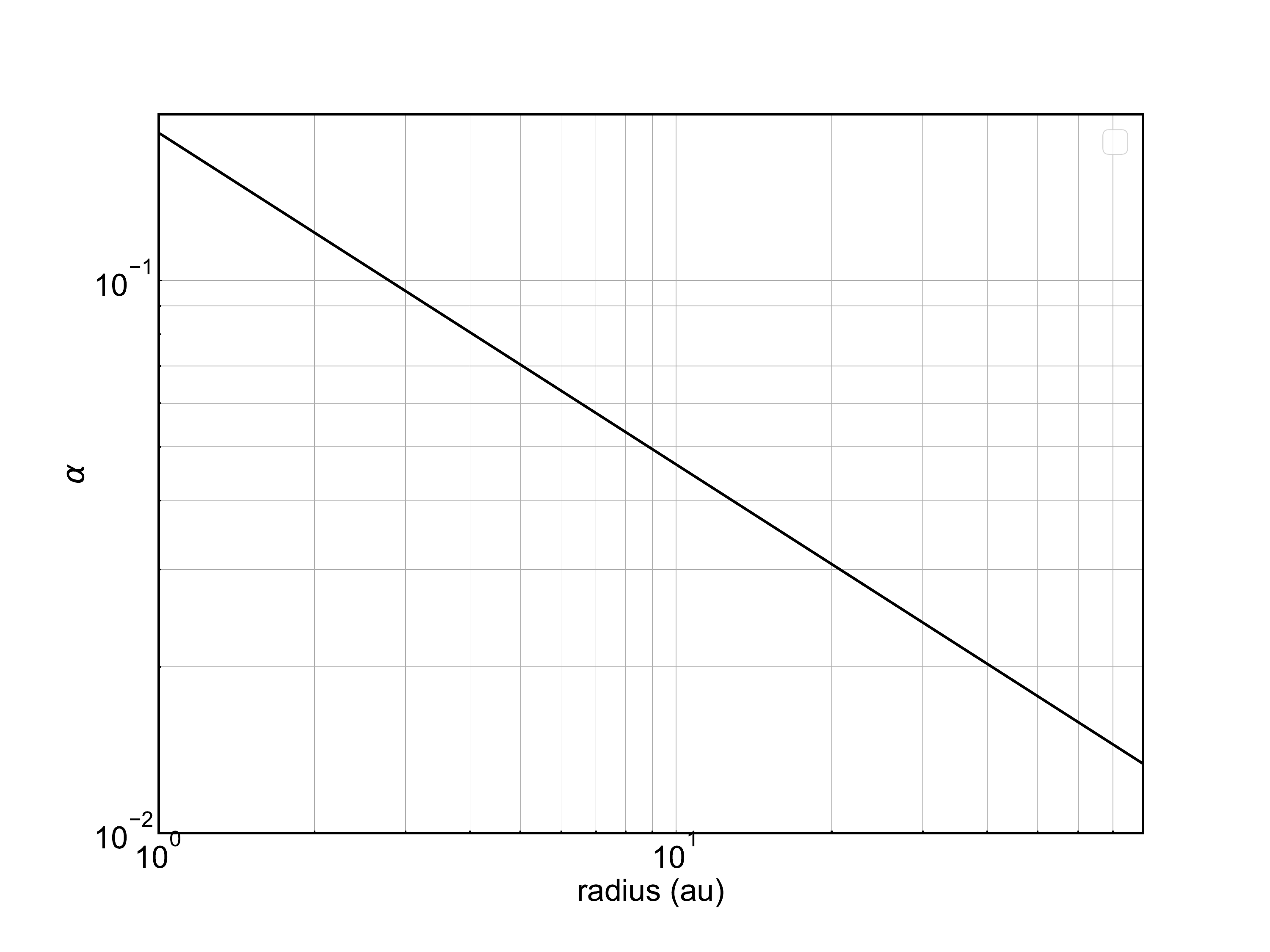}
\caption{Radial Distribution of the $\alpha$ parameter in the disk of the fiducial model.
\label{fig:alpha}}
\end{center}
\end{figure*}

With these gas and dust density distibutions, radiative transfer calculations to produce
images of the 1.3-mm dust-continuum and C$^{18}$O (2--1) emission
are conducted using RADMC-3D \citep{2012Dullemond}.
The RADMC-3D calculations involve three separate steps;
1) thermal Monte Carlo simulations to obtain the spatial dust temperature distribution
self-consistently, radiative transfer calculations to produce
images of 2) the dust-continuum emission and 3) the emission from the
C$^{18}$O (2--1) line. To perform the thermal Monte Carlo simulations,
it is required to specify the properties of the central heating source
($i.e.,$ protostar), number of photons, wavelength-dependent dust opacity,
and if necessary spatial distribution of the internal heating rate.
RADMC-3D requires the flux density and radius to specify
the central protostar in general. For simplicity,
the radiation from the protostar is assumed to be
blackbody radiation from a point source.
In such a case, RADMC-3D requires the blackbody temperature only.
The relation between the blackbody temperature ($\equiv T_{\star}$)
and the protostellar luminosity ($\equiv L_{\star}$) is expressed as
\begin{equation}
T_{\star} = \left(\frac{L_{\star}}{4\pi R_{\star}^2 \sigma_{sb}} \right)^{\frac{1}{4}},
\end{equation}
where $R_{\star}$ and $\sigma_{sb}$ denote the protostellar radius
and Stefan–Boltzmann constant, respectively.
$L_{\star}$ is set to be the bolometric luminosity of IRS~7B-a
\textcolor{black}{(= 5.2 $L_{\odot}$)
\footnote{During our modeling effort 
the source bolometric luminosity derived by
the eDisk team was 5.2 $L_{\odot}$, which was later updated to
5.1 $L_{\odot}$ \citep{2023Ohashi}.
Since the small difference of the luminosity yields
a negligible difference of the stellar radius from 1.0 $R_{\odot}$
to 0.99 $R_{\odot}$ to have the same $T_{\star}$,
in the present paper, we adopt 5.2 $L_{\odot}$ as the canonical
value of the stellar luminosity.}
and $R_{\star}$ = 1 $R_{\odot}$, which gives $T_{\star}$ = 8714 K.}
We note that $L_{\rm bol}$ is not strictly the same as $L_{\star}$.
$L_{\rm bol}$ is derived from the SED fitting
over the NIR to millimeter wavelengths \citep{2023Ohashi}.
The NIR emission originated
from the protostellar sphere should be attenuated by the
surrounding disk, envelope and the ambient gas. On the other hand,
the attenuated NIR radiation should be re-emitted at longer wavelengths.
As a total radiation energy $L_{\rm bol}$ is still a reasonable proxy
of $L_{\star}$.
To estimate the error of $L_{\rm bol}$, we tried a new SED fitting
using only the data points with the highest value at each wavelength
\citep[see Figure 9 in][]{2023Ohashi}.
The derived $L_{\rm bol}$ is 8.3 $L_{\odot}$, $\sim$60$\%$ increase from the
adopted value of $L_{\rm bol}$ = 5.2 $L_{\odot}$.
As shown below,
this highest possible value of $L_{\rm bol}$ is still
too low to reproduce the observed high brightness temperature
of the 1.3-mm dust continuum emission.

Note that with a given $L_{\star}$, $R_{\star}$ and $T_{\star}$ are degenerate, and
larger $R_{\star}$ yields lower $T_{\star}$. $R_{\star}$ = 1 $R_{\odot}$ is likely
a lower limit of the protostellar radius, so $T_{\star}$ is likely
an upper limit.
The background of this parameter choice
is the observed bright 1.3-mm dust-continuum emission.
As we will describe below, we have found that the observed bright 1.3-mm dust-continuum emission
cannot be reproduced with the heating from the central protostar only,
and that internal heating in the disk is likely required.
As RADMC-3D only requires the central blackbody temperature to calculate
the heating from the central protostar, we set this temperature to a high value.
We also attempted even higher luminosities and
thus higher blackbody temperatures in our parameter search.

The dust mass opacity is also important but rather uncertain.
To convert the observed 1.3-mm flux densities to the dust masses,
eDisk papers adopt the Beckwith opacity,
which gives $\kappa_{\rm 1.3 mm}$ = 2.3 cm$^{2}$ g$^{-1}$
with $\beta$ = 1 \citep{Beckwith1990}.
This $\kappa_{\rm 1.3 mm}$ value is similar to
the values inferred in the disks around the Class 0 source HH 212 mms \citep{2021Lin}
and the Class I source IRAS 04302+2247 \citep{2023Lin}.
The Beckwith opacity is similar to the opacity with a maximum grain size of 1 mm
at millimeter wavelenghs \citep{2011Andrews,2018Birnstiel,2023Han}.
Another adopted opacity in our modeling is the so-called DSHARP opacity \citep{2018Birnstiel}.
While a variety of DSHARP opacity tables can be obtained
depending on the assumed dust parameters,
the adopted DSHARP opacity in our models 
is the same as that shown
in the blue curve in Figure 6 by \citet{2018Birnstiel}.
The adopted opacity is an average over grain sizes
from 0.1 $\micron$ to 1 mm with a power-law index of
the grain size distribution of -3.5,
where the grain composition is summarized in Table 1
in \citet{2018Birnstiel}.
In our parameter search the Beckwith and DSHARP opacity were mostly used,
but OH5 and Semenov opacities were also attempted (Figure \ref{fig:kappa}).
OH5 is the dust opacity taken from Table 1 column 5 in \citet{Ossenkopf1994}.
The dust model corresponds to grains that have coagulated at a gas density of 10$^6$ cm$^{-3}$
with thin ice mantles from the initial MRN
distribution \citep{1977ApJ...217..425M,1984Draine}. This has been long used
as a ``standard" opacity table for radiative transfer models and mass estimates
for embedded protostars.
%
Regarding the Semenov opacity, the dust opacity table
from \citet{Semenov2003} for the composite aggregate dust of the normal silicate mineralogy at T$<$155 K is adopted.



In addition to the heating from the central protostar, we also incorporate
the viscous accretion heating in the disk.
The energy dissipation rate in the viscous Keplerian disk can be expressed as
\begin{equation}
q_{heat} = \frac{9}{4} \rho \nu_{\rm vis} \Omega_{\rm K}^2,
\end{equation}
where $\nu_{\rm vis}$ denotes the viscocity \citep{1981Pringle,1998DAlessio}.
Assuming the constant $\nu_{\rm vis}$ along the disk vertical direction and
the steady accretion, the mass accretion rate $\dot{M}$ is
\begin{equation}
\dot{M} = 3 \pi \Sigma \nu_{\rm vis}.
\end{equation}
Substituting eq (18) into eq (17) to remove $\nu_{\rm vis}$ yields
\begin{equation}
\begin{split}
&q_{heat} =\frac{9}{4} \rho \Omega_{\rm K}^2 \frac{\dot{M}}{3 \pi \Sigma}\\
&= \frac{3 \dot{M} \Omega_K^2 \rho(R,z)}{4\pi \Sigma(R)}.
\end{split}
\end{equation}
\textcolor{black}{$\dot{M}$ is set to be
the same as that of the outer protostellar envelope.}
The $\alpha$ parameter can also be derived as
\begin{equation}
\alpha(R) = \frac{\nu_{\rm vis}}{c_s h} = \frac{\dot{M}}{3 \pi \Omega_K(R) h(R)^2 \Sigma(R)}.
\end{equation}
Figure \ref{fig:visheat} shows the spatial distribution of the heating rate
by the viscous accretion heating of the fiducial model.
Figure \ref{fig:alpha} shows the radial profile of
the $\alpha$ parameter in the fiducial model.
The profile becomes a simple power-law with the power-law index of -0.6.
The $\alpha$ value of
$\sim$0.01-0.1 is physically reasonable for a relatively young disk
\citep[e.g.,][]{2022Hu}.


With the above setting, thermal Monte Carlo simulations are performed to
calculate the spatial distributions of the dust temperature.
To increase the computational speed,
\textcolor{black}{the ``modified random walk mode" implemented in RADMC-3D is adopted,
and the computation is performed in parallel with 30 threads.}
Then radiative transfer calculations to generate the model 1.3-mm dust continuum images
are made. For simplicity, we do not incorporate dust scattering in the radiative transfer calculations.
Dust scattering suppresses the observed intensities \citep{2016Yang,Yang2017,2020Yang,2020Lin,2022Lin,2023Lin},
while our model needs to explain the observed high brightness temperature of the 1.3-mm dust-continuum emission. Thus, inclusion of dust scattering does not change our conclusion that a dust disk passively heated by the central protostar solely cannot reproduce the observed brightness temperature.

On the assumption of equal dust and gas temperatures,
radiative transfer calculations to generate the model C$^{18}$O (2--1) image cubes
are also conducted.
\textcolor{black}{The LVG mode of RADMC-3D ($i.e.$, linemode=2) is adopted to
calculate the non-LTE population distributions of C$^{18}$O, on the assumption
of the maximum photon escape length of 10~au.
}
The C$^{18}$O rotational energy levels, transition frequencies, Einstein A-coefficients, and collisional coefficients with the ortho and para H$_2$ are taken from the Leiden Atomic and Molecular Database \citep[LAMDA;][]{Schoier2005}.


After the radiative transfer calculations are completed,
observing simulations
are conducted to directly compare the model and observed images.
The CASA task $ft$ was used to sample the model images with the observed $uv$ coverages of the IRS~7B data,
and then the CASA task $tclean$ was used to make the model images
\textcolor{black}{with the same imaging parameters as those used for the real observational data.}
In the case of the C$^{18}$O model, the sampled model visibility was continuum-subtracted with the CASA task $uvcontsub$, and then
$tclean$ was used to generate the line image cubes.
Details of the imaging of the eDisk data are given by
\citet{2023Ohashi}.

\subsection{Parameter Search} \label{sec:param}

Our model requires a number of parameters and
ALMA observing simulations, and thus substantial computational time.
In addition, there is a possible offset of the peak position
toward the southeast (Figure \ref{fig:dustobs}).
The offset of the strong peak
prevents us from calculating the numerical measure of the goodness
of our axissymmetric model. Therefore, we do not adopt
a numerical procedure to fit the model to the observed image
such as MCMC.
Instead, we searched for model parameters which
decently reproduce the observed images by eye.
Comparisons between the models and observations were
only performed on the image plane. This is because
the image-based comparison is more intuitive,
and the presence of the binary companion (IRS 7B-b)
complicates the modeling of visibilities.
To reduce the dimension of the parameter space,
limited searching parameters,
which we consider directly affect the images
and intensities, are chosen. Those include
the disk mass ($M_{d}$), gas and dust disk radii
($r_{gas}$ and $r_{dust}$), disk flaring index $q$,
mass accretion rate ($\dot{M}$), disk inclination angle ($i$),
protostellar luminosity ($L_{\star}$),
and the dust mass opacity table.
On the other hand, parameters such as the
source distance, disk position angle,
and the coordinates and wavelength of the observations
are known a priori, and those parameters are fixed.
Other fixed parameters are assumed to have canonical values
often used in the literatures, or determined by the small
preparatory parameter search
(such as the protostellar mass $M_{\star}$ = 2.9 $M_{\odot}$).
The fixed parameters of the model are summarized in Table \ref{fixed}.

For the dust opacity table
of Beckwith and DSHARP, we first tried to reproduce the
extent and the aspect ratio of the observed image in the
1.3-mm dust continuum emission, and to adjust $r_{dust}$, $M_{d}$,
and the inclination angle. Next we reproduced the peak intensity
of the continuum emission by adjusting $\dot{M}$ and $M_{d}$.
We then checked that our choise of the power law index
of the radial surface density profile ($p$ = -0.5) gives
a decent radial intensity profile along the major axis.
Next we attempted to reproduce the observed skewed intensity profile
along the minor axis. We found that the disk flaring index $q$ and
$M_{d}$ most affect this. As we will discuss below, $q$ and $M_{d}$
are degenerate to reproduce the asymmetric
intensity profile along the minor axis,
and it is not straightforward to pinpoint the best set of these two
parameters. We adopt $M_{d}$ = 0.41 $M_{\odot}$ and $q$ = 0.3,
both of which appears to be reasonable, as fiducial values.
For the dust opacity table of OH5 and Semenov,
only a small parameter search around the fiducial values were performed.
These additional searches have revealed that
the choice of the dust opacity table does not significantly affect
the model 1.3-mm images.

Finally, with the given model image of the dust-continuum emission,
we tried to reproduce
the observed C$^{18}$O image cube by changing $r_{gas}$. Here, we
attempted several values of the enhanced gas-to-dust mass ratio between
$r_{dust}$ and $r_{gas}$ and determined the fixed value of
the gas-to-dust mass ratio of 5000.
The factor of 50 enhancement of the gas-to-dust mass ratio
was chosen to keep the dust temperature distribution flat. If the amount of
the dust is further reduced, the dust temperature in that region increases.
To verify the larger $r_{gas}$ than $r_{dust}$, we also changed $r_{dust}$
and checked the resultant model C$^{18}$O image cubes.


Our fiducial model is model 61 in Table \ref{search},
where \textcolor{black}{
the disk flaring index $q$ = 0.30, gas+dust disk mass $M_{d}$ = 0.41 $M_{\odot}$,
dust disk radius $r_{dust}$ = 62~au, gas disk radius $r_{gas}$ = 80~au,
mass accretion rate $\dot{M}$ = 1.4 $\times$ 10$^{-6}$ $M_{\odot}~ {\rm yr}^{-1}$,
disk inclination $i$ = - 70$\degr$, the central
stellar luminosity $L_{\star}$ = 5.2 $L_{\odot}$, and
the DSHARP opacity are adopted.}
After the fiducial model parameters have been obtained,
we changed one of the parameters to see the dependence of that
particular parameter,
and those parameters include $L_{\star}$, $M_{d}$, $q$, and
$\dot{M}$, i.e., inclusion or exclusion of the viscous accretion heating.
All the varied model parameters are listed in Table \ref{search} in Appendix.


\begin{deluxetable*}{ll}
\tablecaption{\textcolor{black}{
Fixed Parameters for the IRS 7B-a modeling} \label{fixed}}
\tablewidth{0pt}
\tablehead{\colhead{Parameter} &\colhead{Value}}
\startdata
Radial Range &1 au --20000 au\\
Distance & 152 pc \\
Mass of the Protostar $M_*$ &2.9 $M_{\odot}$\\
Radius of the Protostar $R_*$ &1.0 $R_{\odot}$\\
Disk Position Angle $\theta_d$ &115$\degr$ \\
Power Law Index of the Disk Surface Density $p$ & - 0.5 \\
Temperature at 1 au &400 K \\
Envelope Flattening Factor $\eta_f$ & 2.0 \\
Radial Range of the Static Cocoon & 10000 au -- 20000 au\\
Gas Density of the Static Cocoon $n^{coc}_{\rm H_2}$ & 10$^4$ cm$^{-3}$ \\
Turbulent Velocity Dispersion in the Disk $\sigma_{disk}$ & 0.0 km s$^{-1}$\\
Turbulent Velocity Dispersion in the Envelope and Cocoon $\sigma_{env}$ & 0.2 km s$^{-1}$\\
Dust Continuum Wavelength & 1.34 mm\\
C$^{18}$O Abundance $X^{can}_{\rm C^{18}O}$ &1.76$\times$10$^{-7}$ \\
Gas to Dust Mass Ratio ($r < r_{dust}$, $r_{gas} <r$) & 100 \\
Gas to Dust Mass Ratio ($r_{dust} <r < r_{gas}$) & 5000 \\
Ortho-to-Para Ratio of H$_2$ & 3.0 \\
Photon Escaping Spatial Scale & 10 au \\
Number of photons & 1.5$\times$10$^6$\\
\enddata
\end{deluxetable*}

\section{Modeling of the 1.3-mm Dust-Continuum Image}
\label{sec:dustmodel}

\subsection{1.3-mm Dust-Continuum Intensity} \label{sec:intensity}

\begin{figure*}[ht!]
\begin{center}
\includegraphics[width=150mm, angle=0]{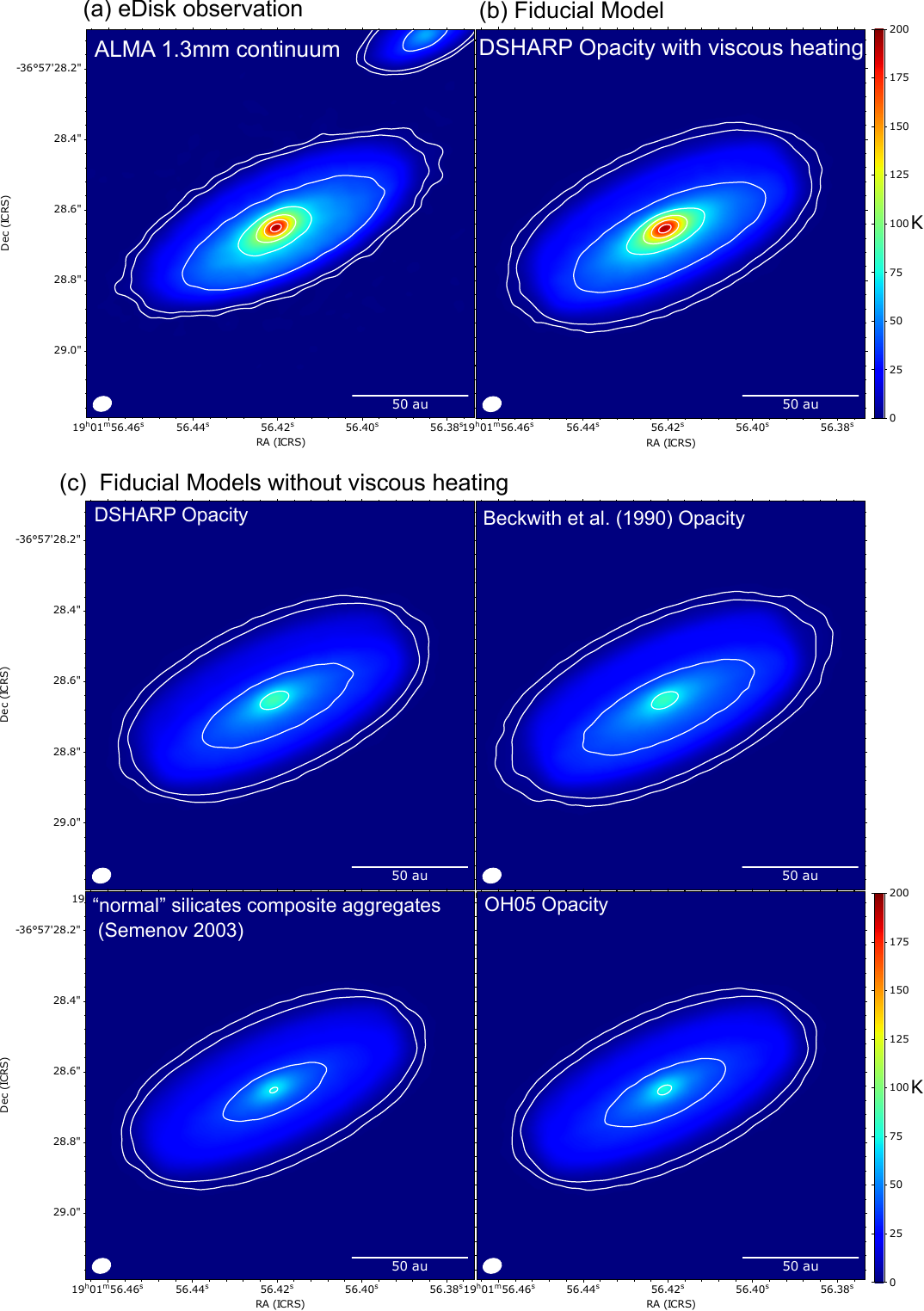}
\caption{Comparison of the observed 1.3-mm dust-continuum image (a)
to the model images with (b) or without the viscous accretion heating (c).
Contour levels are the same as those in Figure \ref{fig:dustobs}.
The model image with the viscous accretion heating adopts
the DSHARP opacity and the mass accretion rate of
\textcolor{black}{$\dot{M} = 1.4 \times 10^{-6}~M_{\odot}~ {\rm yr}^{-1}$,}
while the model images without the heating adopt four different dust opacities as labeled.
\textcolor{black}{Panel (b) corresponds to model 61
in Table \ref{search}, and the four models in Panel (c) correspond
to models 66, 68, 69, and 70, respectively.}
\label{fig:obsmodel}}
\end{center}
\end{figure*}

\begin{figure*}[ht!]
\begin{center}
\includegraphics[width=170mm, angle=0]{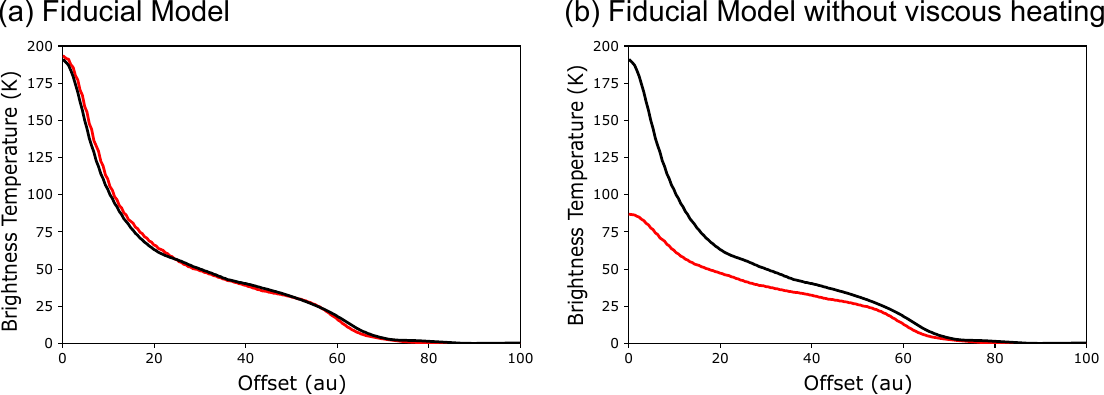}
\caption{Comparison of the observed and model intensity profiles
of the 1.3-mm dust-continuum emission. Black curves
in both panels (a) and (b) show the observed intensity profile averaged
in the northwest and southeast directions along the major axis.
Red curves in panels (a) and (b) present the model intensity profiles
with (model 61 in Table \ref{search}) and without (model 66)
the viscous accretion heating, respectively.
\label{fig:dustprof}}
\end{center}
\end{figure*}

\begin{figure*}[ht!]
\begin{center}
\includegraphics[width=170mm, angle=0]{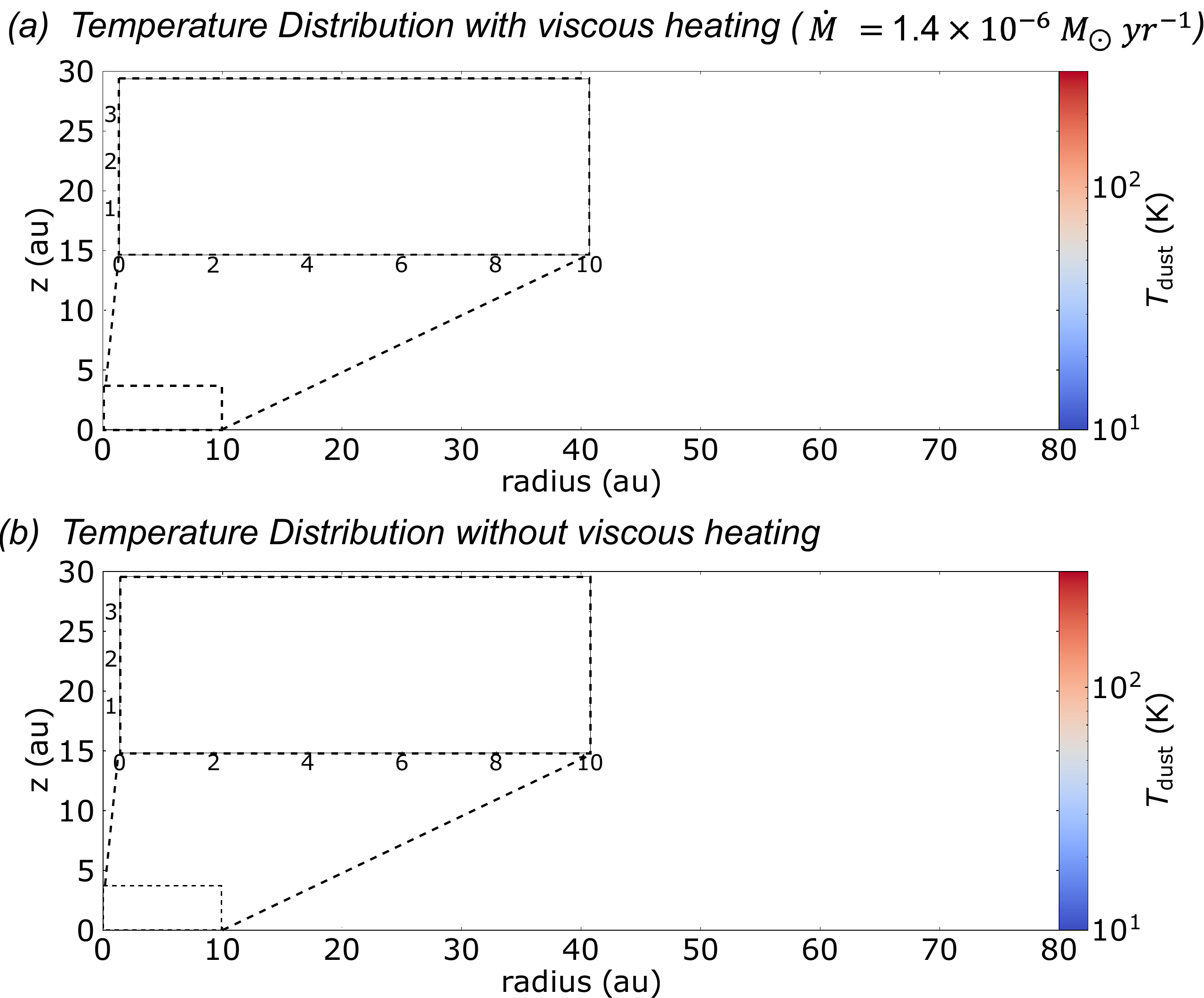}
\caption{Distribution of the dust temperature in the disk region
with \textcolor{black}{(a; model 61 in Table \ref{search})}
and without the viscous accretion heating
\textcolor{black}{(b; model 66).}
\textcolor{black}{Contour levels are 25, 50, 75, 100, 200, 300, 400, and 500 K.}
\label{fig:tempdist}}
\end{center}
\end{figure*}

\begin{figure*}[ht!]
\begin{center}
\includegraphics[width=180mm, angle=0]{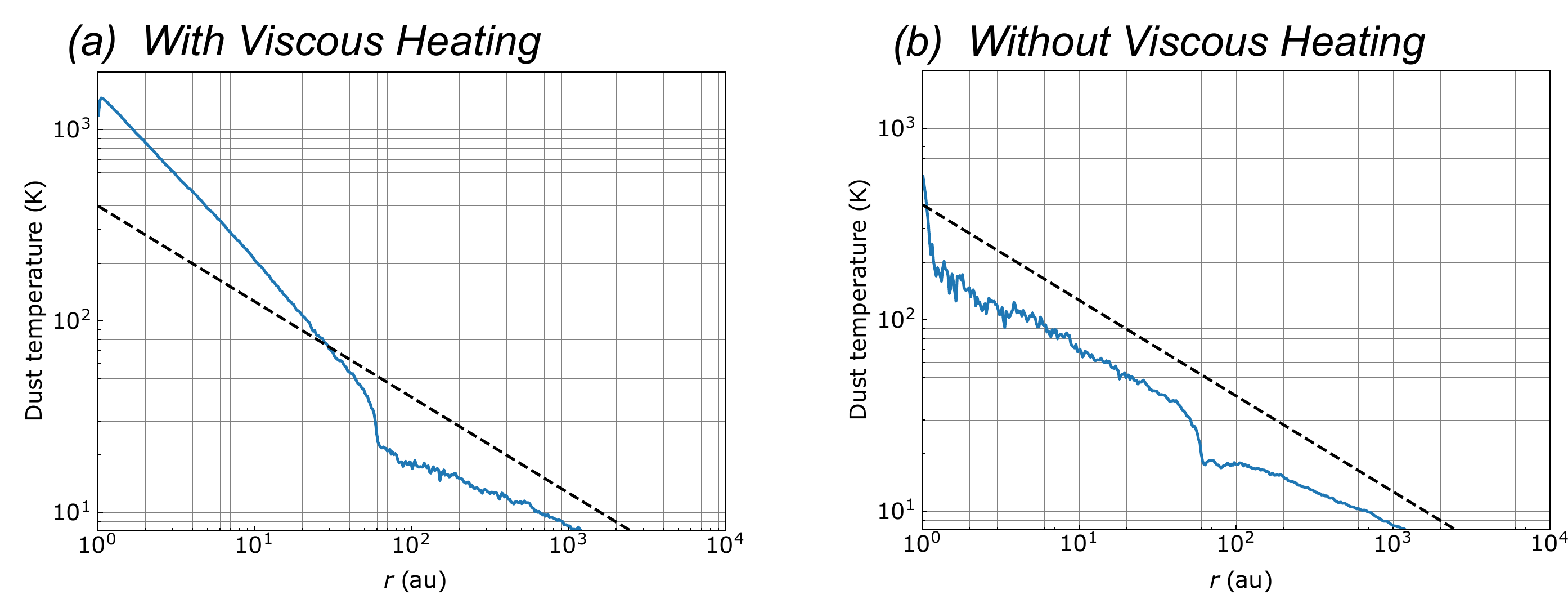}
\caption{Radial temperature profiles (blue lines) in the disk midplane with \textcolor{black}{(a; model 61 in Table \ref{search})}
and without
the viscous accretion heating \textcolor{black}{(b; model 66).}
Dashed lines show the temperature profile
of 400 (K) $\times (r/1~{\rm au})^{-0.5}$.
\label{fig:tempprofile}}
\end{center}
\end{figure*}


We first aim to reproduce the observed bright 1.3-mm intensity 
in the protostellar disk around IRS~7B-a.
Figure \ref{fig:obsmodel} shows the model 1.3-mm dust-continuum images
with and without the viscous accretion heating using the
DSHARP opacity, and those without the viscous heating
using the other three opacities.
Comparison of the observed and model intensity profiles
along the major axis with the DSHARP opacity is shown in
Figure \ref{fig:dustprof}.
Without internal heating,
the peak intensity of the 1.3-mm dust-continuum emission is only $\lesssim$ half of
the observed intensity, regardless of the adopted opacity
(Figure \ref{fig:obsmodel}c and Figure \ref{fig:dustprof}b).
Incorporation of the viscous accretion heating raises
the model intensities and reproduces the observed intensities
(Figure \ref{fig:obsmodel}b and Figure \ref{fig:dustprof}a). This is the case for all the opacity tables.

We tried to reproduce
the observed 1.3-mm intensity with only the passive heating from the
central protostar,
\textcolor{black}{adopting very high protostellar luminosities of
$L_{\star}$ = 20 $L_{\odot}$ (Model 27 in Table \ref{search}),
26 $L_{\odot}$ (Model 49), and 52 $L_{\odot}$ (Model 60).
We found that
the central luminosity of 52 $L_{\odot}$ is required to
reproduce the disk brightness.}
This luminosity is, however, one order of magnitude higher
than the measured bolometric luminosity.
Even if the uncertainty of the $L_{\rm bol}$ measurement
is taken into account, this luminosity
is still a factor of $\sim$seven higher than
the possible highest value of $L_{\rm bol}$ (see Section \ref{sec:RADMC-3D}).
We also changed the disk mass and dust mass opacity
and checked how the model 1.3-mm dust continuum intensity
changes with these parameters (see Table \ref{search}).
Note that the disk mass and the dust mass opacity
at the observed wavelength $\kappa_{\rm 1.3mm}$ are degenerate
in controlling the 1.3-mm intensity.
We found that increasing $M_d$ to 0.55 $M_{\odot}$
does not increase the model 1.3-mm intensity (model 34).
Adopting the different dust opacities seldom changes the 1.3-mm
intensity (Figure \ref{fig:obsmodel}c).
This indicates that the
model 1.3-mm dust-continuum emission in the protostellar disk
is optically thick.
Thus, the only remaining way to raise the continuum
intensity is to increase the dust temperature in the disk.
Figures \ref{fig:tempdist} and \ref{fig:tempprofile} show the
temperature distributions of the models with and without viscous heating,
which correspond to Figure \ref{fig:obsmodel}b and c with the DSHARP opacity.
It is clear
that incorporation of the viscous accretion heating raises the midplane dust
temperature in the inner $\lesssim$50~au,
which enhances the dust emission intensities.

We conclude that the observed bright 1.3-mm intensity in the protostellar disk around
IRS~7B-a cannot be reproduced with passive heating from the central protostar only. In other words, the protostellar disk
is ``self-luminous''.


\subsection{Asymmetry along the Minor Axis}
\label{sec:asymmetry}

\begin{figure*}[ht!]
\begin{center}
\includegraphics[width=185mm, angle=0]{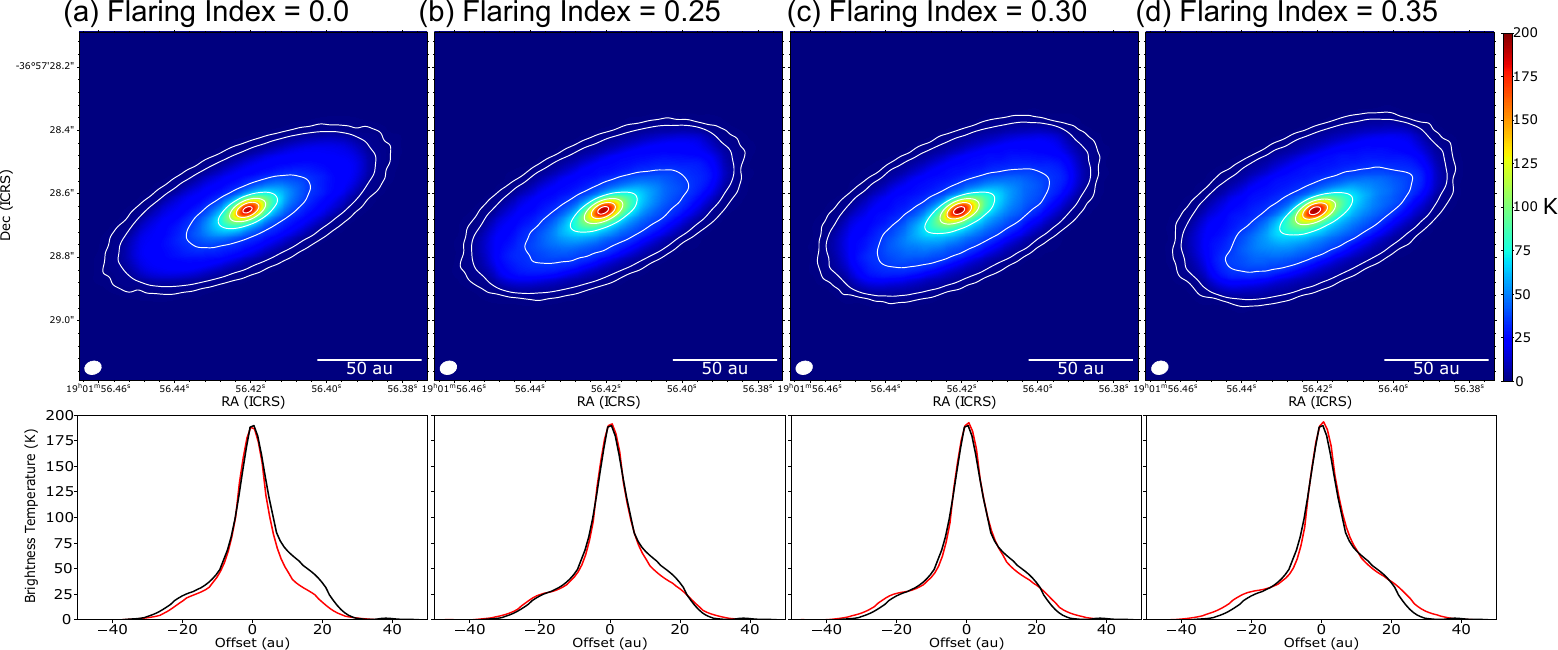}
\caption{Upper panels: Model 1.3-mm dust-continuum images
with different flaring indices of the dust distribution as labeled.
\textcolor{black}{Contour levels
are the same as those in Figure \ref{fig:dustobs}.}
Lower panels: Relevant intensity distributions along the minor axis (red curves)
to be compared to the observed intensity profile (black).
\textcolor{black}{Panels (a), (b), (c), and (d) correspond to models
4, 28, 61 (Fiducial), and 85, respectively, in Table \ref{search}.}
\label{fig:minorshift}}
\end{center}
\end{figure*}

\begin{figure*}[ht!]
\begin{center}
\includegraphics[width=170mm, angle=0]{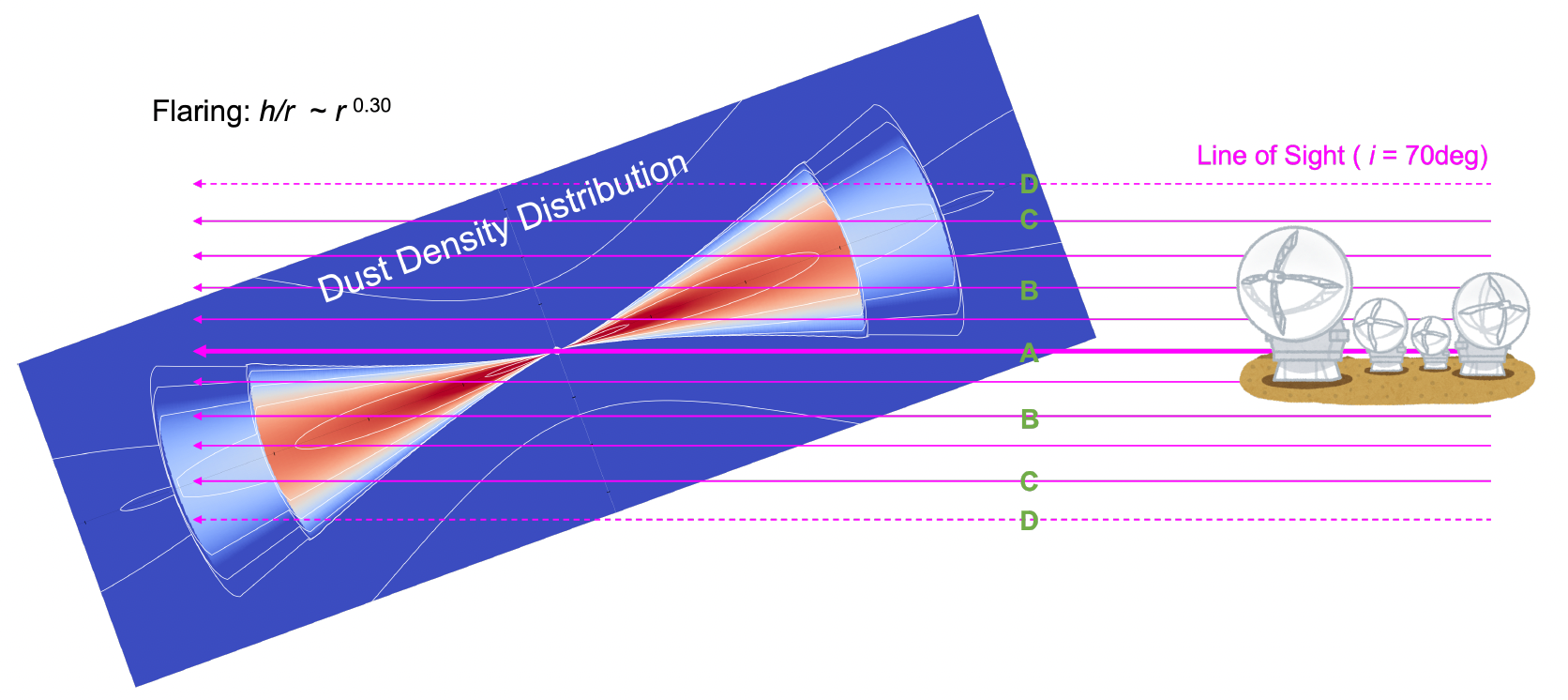}
\caption{Schematic picture to interpret the skewed distribution
of the 1.3-mm dust-continuum emission along the minor axis.
The dust distribution in the model disk, inclined by 70$\degr$
to the line of sight of the observer, is shown. Line of sights
toward the center (A) and the offset positions (B,C,D) are drawn.
\textcolor{black}{Contour levels and color scale are the same
as those in Figure \ref{fig:modelconf}c}.
\label{fig:schemedust}}
\end{center}
\end{figure*}

As well as the 1.3-mm dust-continuum intensity, the observed
skewed distribution along the minor axis should be reproduced.
Figure \ref{fig:minorshift} shows model images (top)
and intensity
profiles along the disk minor axis (bottom) with different disk flaring indices $q$.
\textcolor{black}{All other model parameters are set to the fiducial values.}
The flaring index $q$ is defined in eq. (5).
$q$ = 0 indicates that the dust
scale height increases linearly with $r$.
$q$ = 0.25 corresponds to the disk midplane temperature being $\propto r^{-0.5}$ and the disk being in
the vertical hydro-equilibrium. In the case of $q$ = 0,
the model image exhibits a fairly symmetric feature
(Figure \ref{fig:minorshift}a). The model intensity
profile along the minor axis (red curve) shows little
deviation from the symmetry. On the other hand,
the observed intensity profile (black) shows a clear asymmetry,
and there is a shallow shoulder to the southwest. If a steeper
flaring index is adopted, the model images
and intensity profiles exhibit more skewness along the minor axis.
$q$ = 0.25 reproduces the northeastern part of the observed intensity profile fairly well, while that index appears to be
not enough to reproduce the southwestern shoulder
(Figure \ref{fig:minorshift}b).
In the case of $q$ = 0.30 the southwestern shoulder appears
to be better reproduced, while there is a small deviation 
to the northeastern part (Figure \ref{fig:minorshift}c).
If an even steeper flaring index
is adopted (Figure \ref{fig:minorshift}d), the southwestern
shoulder becomes too shallow. 



\textcolor{black}{
We found that if archetypal disk models with a flaring of $\frac{h}{r} \propto r^{0.25}$ and a mass of
$\lesssim$ 0.1 $M_{\star}$ are adopted,
the model image exhibits a fairly symmetric
intensity profile along the minor axis (models 12-18).
It is necessary to raise the disk mass to
$\sim$0.41 $M_{\odot}$ $\approx$0.14~$M_{\star}$ to reproduce the
observed asymmetry along the minor axis. On the other hand,
a very high disk flaring $q \gtrsim$ 0.5 is required to reproduce
the asymmetry with a low disk mass (models 96 -- 99).
We found that either a higher flaring index or a higher disk mass is required
and that the flaring index and the disk mass are degenerate,
to produce the skewed emission distribution.}
As fiducial values $q$ = 0.3 and $M_{\rm disk}$ = 0.41 $M_{\odot}$ are adopted.

Figure \ref{fig:schemedust} presents a schematic figure to explain
the cause of the asymmetry along the minor axis.
We consider the line of sights (LOSs) B and C offset from the LOS
toward the disk center (LOS A). Even though the amount of
the offset of those LOSs
from the center are the same on the upper
and lower sides in Figure \ref{fig:schemedust},
those on the lower side directly see the hot disk surface
if the dust-continuum emission there is optically thick. 
On the other hand, on the upper side,
the LOSs encounter the colder midplane regions.
Thus, the overall emission distribution
should be skewed to the lower side.
If the LOSs are far from the disk center (LOS D),
those portions become optically thin. Thus, at larger radii, the asymmetric distribution becomes less significant. These portions correspond to the elliptical region
as denoted by a dashed line in Figure \ref{fig:dustobs}.
If the disk is more inclined or more flared, the LOS toward the disk center (LOS A) passes
through a portion of the flared disk in the near side.
The near-side portion of the flared disk affects
the observed intensity at the central LOS, and the central position can be dimmer.
The extreme case is the perfectly edge-on disk in HH 212 mms,
which produces a dark lane at the disk midplane and
a morphology shaped as a ``space hamburger'' \citep{2017Lee}. 

Thus, both disk flaring and optically thickness of the dust emission are required
to produce the skewed emission distribution.
If the disk is perfectly geometrically thin, or the dust emission
is perfectly optically thin, the emission distribution
should always be symmetric. The degeneracy between
the flaring index and disk mass found by our modeling
is naturally explained since more flared disks
require less optical depth, and less flared disks require more optical depth.
On the other hand, disk stability with a given disk mass can be evaluated
with the Toomre $Q$ value as:
\begin{equation}
Q = \frac{c_s \Omega_{\rm K}}{\pi G \Sigma},
\end{equation}
where the formulae of $c_s$, $\Omega_{\rm K}$, and $\Sigma$
are given in eqs (3), (4), (6), respectively. Note that
the $Q$ value depends on $R$.
Adopting the fiducial disk mass of 0.41 $M_{\odot}$ and
the dust temperature shown in Figure \ref{fig:tempprofile}a
to calculate $c_s$, the Toomre $Q$ values are calculated to be
$\sim$11.2, 2.2, and 1.0 at $R$ = 10, 30, and 50~au, respectively.
Thus, the inner region is gravitationally stable, but the outer region
can be unstable.
In combination with an approapriate value of the disk flaring index
of $\sim$0.3, the fiducial value of the disk mass is thus reasonable.


\section{Modeling of the C$^{18}$O Image Cube} \label{sec:c18o}

\begin{figure*}[ht!]
\begin{center}
\includegraphics[width=185mm, angle=0]{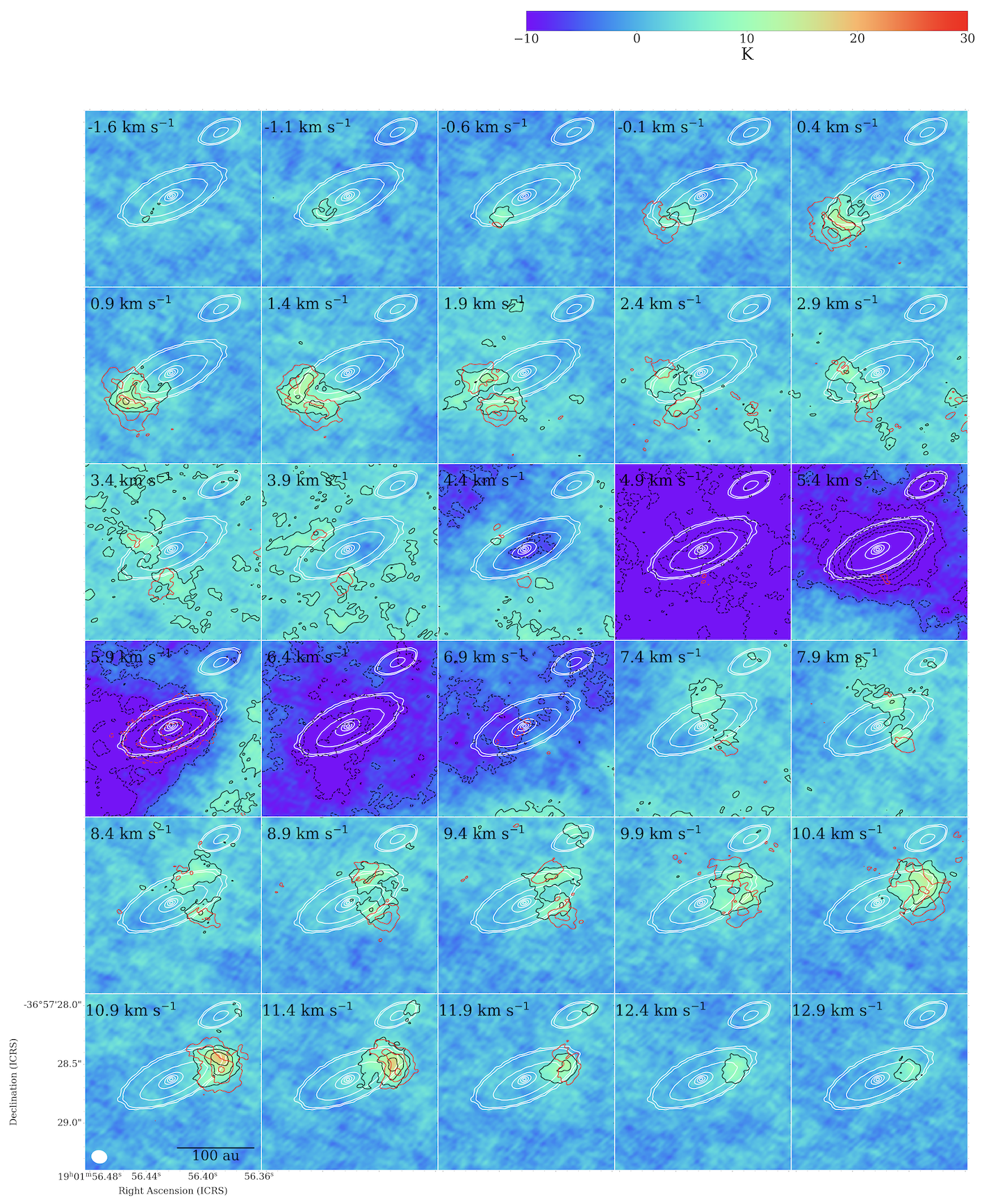}
\caption{Comparison of the observed (black contours and colors) and model
C$^{18}$O (2--1) velocity channel maps (red) in IRS~7B-a
\textcolor{black}{(model 61 in Table \ref{search}).
Contour levels of the C$^{18}$O emission
start from -18$\sigma$ in steps of 3$\sigma$ (1$\sigma$ = 1.72 K).}
White contours show the map of the observed 1.3-mm
dust-continuum emission, where the \textcolor{black}{contour levels are same as
those in Figure \ref{fig:dustobs}.}
\label{fig:c18och}}
\end{center}
\end{figure*}

\begin{figure*}[ht!]
\begin{center}
\includegraphics[width=170mm, angle=0]{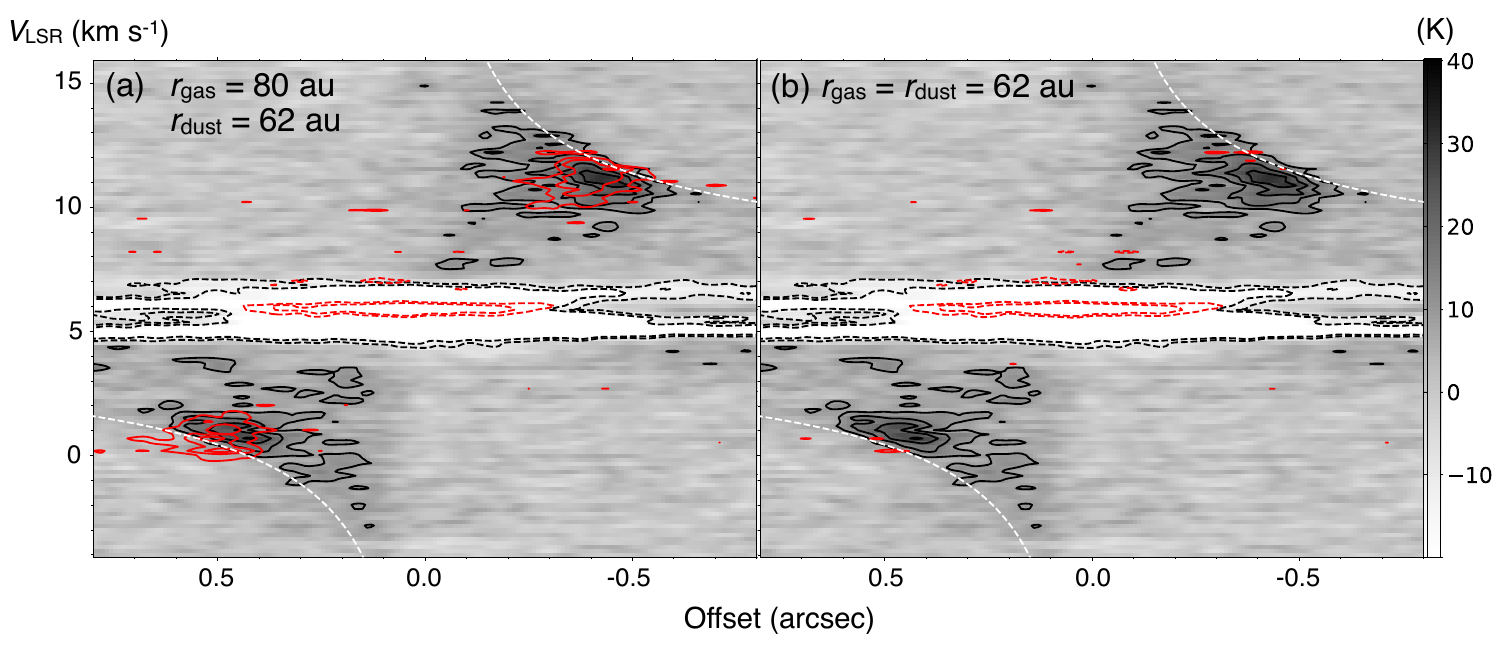}
\caption{Comparison between the observed (black contours
and grayscale) and model P-V diagrams (red contours) of the C$^{18}$O (2--1) emission
along the major axis \textcolor{black}{(P.A. = 115$\degr$).}
Panel (a) shows the model P-V diagram with a dust radius of 62~au
and a gas radius of 80~au including the noise
(model 100 in Table \ref{search}),
while Panel (b) with the gas and dust radius of 62~au
including the noise (model 101).
\textcolor{black}{Contour levels are -5$\sigma$, -3$\sigma$,
3$\sigma$, 5$\sigma$, 7$\sigma$, 9$\sigma$, and 11$\sigma$ (1$\sigma$ = 2.98 K).}
White dashed curves denote the Keplerian rotation curve with
a central protostellar mass of 2.9 $M_{\odot}$ and a disk
inclination angle $i$ = 70$\degr$.
\label{fig:c18opv}}
\end{center}
\end{figure*}


After reproducing the observed 1.3-mm dust-continuum image,
we seek to reproduce the observed intensities and spatial and velocity distributions of the C$^{18}$O (2--1) emission next.
The C$^{18}$O (2--1) emission has been adopted as a good
probe of protostellar disks and envelopes
\citep[e.g.,][]{2014Ohashi,2014Yen,2015Aso}.
Black contours and colors in Figure \ref{fig:c18och} show
the observed velocity channel maps of the C$^{18}$O (2--1) emission in IRS~7B-a \citep{2023Ohashi}. While the C$^{18}$O emission
exhibits a velocity gradient along the disk major axis and a
signature of Keplerian rotation,
the molecular emission is detected only at the outskirts
of the dust disk (white contours in Figure \ref{fig:c18och}).
This is because the molecular emission is buried in the bright
dust-continuum emission (see section \ref{subsec:molline} for
the mechanism of the absence of the molecular emission toward the dust-continuum
emission).

In our modeling of the C$^{18}$O emission
we first adopted the same radius of the gas disk
as that of the dust disk
\textcolor{black}{(e.g., $r_{gas}$=$r_{dust}$=62~au in model 74
and $r_{gas}$=$r_{dust}$=80~au in model 76; see Table \ref{search}).}
In these cases, however,
the C$^{18}$O emission is significantly suppressed, and the
observed C$^{18}$O emission cannot be reproduced.
We found that a larger radius of
the gas disk is required to reproduce the observed intensities
as well as the spatial and velocity distributions of the
C$^{18}$O emission (fiducial model 61).
As both the $r_{gas}$=$r_{dust}$=62~au and $r_{gas}$=$r_{dust}$=80~au
cases yield the significantly suppressed C$^{18}$O emission,
the difference
of the gas and dust radii is unlikely to be due to the effect of the noise
in the continuum data.

\textcolor{black}{To ensure the difference of the gas and
dust radii,
the CASA observing simulator including the noise
is applied to the calculated C$^{18}$O model image cubes.
Note that the signal-to-noise ratio of the continuum emission
is higher than or comparable to that of the C$^{18}$O emission
in the C$^{18}$O emission region (see Figure \ref{fig:c18och}).
Two model image cubes, one
with $r_{gas}$=80~au and $r_{dust}$=62~au (fiducial)
and the other with $r_{gas}$=$r_{dust}$=62~au are adopted
for the CASA simulations. Note that these two models do not adopt
the CASA task $ft$ but $simobserve$ to include the noise
(models 100 and 101, respectively).}
The model velocity channel maps of the
C$^{18}$O emission with the gas disk radius
\textcolor{black}{of 80~au (model 100)} are overlaid
in Figure \ref{fig:c18och} (red contours). The overall locations
of the C$^{18}$O emission, shapes, and intensities
are broadly reproduced
with the model. Figure \ref{fig:c18opv}a compares
the observed and model PV diagrams along the major axis of the
dust disk. The observed emission location in the PV diagram is
reproduced with the model. In contrast, the model with
the gas radius identical to the dust radius (model 101)
cannot reproduce the
observed intensities (Figure \ref{fig:c18opv}b).
These results suggest that the radius of the gas disk is larger
than that of the dust disk in IRS~7B-a.

\section{Discussion} \label{sec:discussion}
\subsection{Self-Luminous Protostellar Disks} \label{subsec:hightd}

Our modeling and parameter search have found that
models of dusty disks only passively heated by the
central protostar cannot reproduce the observed bright 1.3-mm dust-continuum intensities. 
The observed peak brightness temperature is as high as $\sim$195~K.
To reproduce such a high dust brightness temperature, 
internal heating by the viscous accretion, along with the
optically thick dust emission, is incorporated.
Then, with a typical mass accretion rate
of $\sim$a few $\times$10$^{-6}$ $M_{\odot}~yr^{-1}$
the observed brightness temperature in the disk can be reproduced.
In other words, the inner part of the IRS~7B-a disk is self-luminous,
with most of the heating coming from the release of gravitational
energy as the material in the disk spirals inward gradually
as it loses angular momentum.

Such a high peak brightness temperature of the 1.3-mm dust
continuum emission in IRS~7B-a is not unique
but seen in several other eDisk targets, such as
IRAS 04166+2706 (133 K),
BHR 71 IRS1 (170 K), Oph IRS 63 (170 K), R CrA IRAS 32A (153 K),
and TMC-1A (187 K) \citep{2023Ohashi}. While a modeling
effort tailored for these targets is required, these high
brightness temperatures likely suggest that
these protostellar disks are also self-luminous.
In contrast, among the 20 sources in the DSHARP sample,
no source shows the peak brightness temperature of the 1.3-mm
dust-continuum emission higher than
150 K, and most sources show the peak brightness temperatures
lower than $\sim$60 K \citep{2018Andrews}.
Peak 1.3-mm brightness
temperatures of edge-on Class II disks are even lower
\citep[5-10 K;][]{2020Villenave}. These results suggest
a systematic difference of the physical conditions between
the protostellar and protoplanetary disks.

\citet{2023MNRAS.tmp.3602A} constructed detailed models of the dust disk
around V883 Ori, a Class I FU-Ori object, which has been studied as an intriguing
astrochemical laboratory \citep{2018vant,2023Tobin}. Their model also demonstrates
that the viscous accretion heating is required to reproduce the observed high
($\sim$500 K) brightness temperature of the 1.3-mm dust-continuum emission.
The presence of viscous accretion heating in protostellar disks
has also been suggested in other variable sources
\citep{2020MNRAS.495.3614Contreras,2021Yoon} and
a massive protostar GGD 27-MM1 \citep{2020Anez}.
While IRS 7B appears to be more quiescent than V883 Ori, our modeling also proves
that even in the quiescent phase the viscous accretion heating is required to
reproduce the dust brightness.


\subsection{Coupling between Dust and Molecular-Line Emission}
\label{subsec:molline}

In this subsection, we argue that
the observed high brightness temperatures of the dust emission
affect the interpretation of the
molecular line images. We here consider two simple cases
\citep[see][]{2021Bosman}.
One is that the dust and gas are located in separate layers
(e.g., disk and envelope, respectively),
and that the optically-thick dust emission resides
behind the foreground molecular emission.
In such a case, the observed brightness temperature of the molecular line ($\equiv T^{mol}_{B}$) is expressed as,
\begin{equation}
T^{mol}_{B} = (J_{\nu}(T_{\rm ex})-J_{\nu}(T_{\rm dust}))(1-\exp(-\tau_{\nu,\rm gas})),
\end{equation}
where
\begin{equation}
J_{\nu}(T) = \frac{\frac{h_P\nu}{k_{\rm B}}}{1-\exp(\frac{h_P\nu}{k_{\rm B} T})}.
\end{equation}
In the above expressions, $\nu$ is the line frequency, $h_P$ is
Planck constant,
$T_{\rm ex}$ is the excitation temperature of the molecular line, and
$T_{\rm dust}$ is the dust temperature. $\tau_{\nu,\rm gas}$ is the
optical depth of the molecular line.
If $T_{\rm ex}$ is close to
$T_{\rm dust}$, $T^{mol}_{B}$ is close to zero,
no matter what the value of $\tau_{\nu,\rm gas}$ or the molecular
column density is. If $T_{\rm ex}$ is lower than $T_{\rm dust}$,
the line shows an absorption.

The other case is that both dust and gas are in the same space
with the same volume.
$T^{mol}_{B}$ is expressed as:
\begin{equation}
\begin{split}
&T^{mol}_{B} =J_{\nu}(T)(1-\exp(-(\tau_{\nu,\rm dust}+\tau_{\nu,\rm gas}))\\
& -J_{\nu}(T)(1-\exp(-\tau_{\nu,\rm dust})) \\
&= J_{\nu}(T) \exp(-\tau_{\nu,\rm dust}) (1 -\exp(-\tau_{\nu,\rm gas})).
\end{split}
\end{equation}
Here we assume $T_{\rm ex}$=$T_{\rm dust}$=$T$.
In this case, if $\tau_{\nu,\rm dust} >> \tau_{\nu,\rm gas}$, $T^{mol}_{B}$ is close to zero. Note that the absolute value of
the dust optical depth does not matter.
In the case of $\tau_{\nu,\rm dust} >> \tau_{\nu,\rm gas}$,
$T^{mol}_{B}$ is always very low even if $\tau_{\nu,\rm dust} < 1$.

The depression or even absorption of the molecular line emission
toward the continuum emission
is commonly seen in ALMA observations of protostellar sources, including our eDisk observations \citep{2023Hoff,2023Linedisk}. These phenomena are often referred to as
``continuum over-subtraction.'' From the considerations above,
a possible reason for the absence of the molecular line emission toward the continuum emission is either line excitation
temperatures close to or lower than the dust temperatures,
or the dust optical depth higher than the line optical depth,
or combination of both.
Both high dust temperature and high dust optical depth produce high observed dust brightness temperatures.
In the case of protostellar sources, molecular emission are originated
not only from the protostellar disk but also from the protostellar envelope.
At lower velocities, the molecular emission from the disk is blended
with that from the colder envelope. Such envelope blending
induces suppression or absorption of the molecular line toward the bright dust disk.
The observed velocity channel maps indeed show that
around the systemic velocity
($V_{\rm LSR}$ = 4.90-6.40 km s$^{-1}$), there is no observed
line emission but a strong absorption feature 
(Figure \ref{fig:c18och}). The C$^{18}$O model
does not predict any emission in this velocity range. While
the line width of the absorption feature is narrower,
the C$^{18}$O model also reproduces the absorption feature at
$V_{\rm LSR}$ = 5.9 km s$^{-1}$ as shown in dashed red contours
in Figure \ref{fig:c18och}.

To trace gas structures, kinematics and chemistry in the inner parts of
dusty disks, molecular lines
which are optically thicker than the dust emission are required.
On the other hand, if the optical depth of that molecular line is too high,
the molecular line cannot trace the disk internal structures
\citep{2018Liu}. In the case of embedded sources, optically-thick
molecular lines trace only the cold outer envelopes, which results in
the absorption againt the bright continuum emission originated from the disk.
It is not straightforward to identify appropriate molecular
tracers for protostellar disks.
Different degrees of the missing flux between the continuum and line images
in interferometric observations also complicate the interpretation of the observed images.
If the original line + continuum image
is totally uniform because of the optically-thick line emission, the
line + continuum image after the interferometric observations is severely suppressed.
After the continuum subtraction, the line only image shows the negative signal
at the continuum position.

It is important to recall that our model assumes
a constant C$^{18}$O abundance and gas-to-dust mass ratio in the
disk and envelope, except for the outer dust-free region in the disk
(Figure \ref{fig:modelconf}).
Our modeling demonstrates that even in such a case
the C$^{18}$O emission is suppressed in the disk.
The observed image could be misunderstood as the CO depletion in the disk.
The same warning applies to the analysis of other lines.
Detailed radiative transfer modeling coupled with the interferometric observing simulations is 
required to interpret the interferometric images correctly and to quantitatively study the physical and chemical conditions of protostellar disks.


\subsection{Dust Distribution in the Protostellar Disk}
\label{subsec:dustdist}

Our modeling also found that the observed skewness of the 1.3-mm dust-continuum emission
along the disk minor axis can be reproduced with the flared disk model
and that the dust emission is optically thick.
Dust in the protostellar disk is yet to be settled.
Such asymmetric distributions of the 1.3-mm
dust continuum emission along the disk
minor axes are also observed in
other eDisk targets, including CB 68 \citep{2023Kido}, L1527 IRS \citep{2023Hoff},
IRAS 04302+2247 \citep{2023Linedisk}, and GSS30 IRS 3 \citep{2023Santa},
as well as the space hamburger in HH212 mms \citep{2017Lee}.
These results suggest that at least in several of the protostellar disks
the dust vertical distributions are flared \citep{2023Ohashi}.

The above result is in sharp contrast to the observed geometrically
thin dust distributions in protoplanetary disks around Class II sources.
Concentric ring/gap features are commonly seen
in such Class II disks. If the dust distributions 
are flared vertically and the disk rotational axes are inclined with respect to
the LOS, such ring/gap features can easily be hidden by
the geometrical effect. Thus, the dust is already settled
in such more evolved disks \citep{2016Pinte}.
The 1.3-mm dust-continuum images of highly
inclined ($i~\gtrsim70\degr$) Class II disks also show
ring/gap features, which strongly implies that the dust
distribution is geometrically thin
\citep{2020Villenave,2022Villenave}.
Among the eDisk targets, two of the most evolved sources,
L1489 IRS and Oph IRS63, show
signatures of concentric rings/gaps in the disks
\citep{2023Ohashi,2023Yamato,2023Flores}.
These disks do not show asymmetry along the disk minor axes.
Dust settling and formation of ring/gap features in the disks
are thought to likely proceed between the Class I to Class II stages,
as also discussed in \citet{2023Ohashi}.

On the other hand, our combined dust continuum and
C$^{18}$O modeling has also revealed that the radius of
the gas disk ($\sim$80~au)
should be larger than that of the dust disk ($\sim$60~au)
around the Class I protostar IRS~7B-a. 
In more-evolved, protoplanatary disks around Class II sources,
observationally-measured radii of CO disks ($e.g.$, curves of growth)
are systematically larger than the corresponding dust radii,
and the gas radii are 2-2.5 times larger than the dust radii
\citep{Ansdel2018,2020Andrews}.
The effect of the optical depth and radial variation of dust $\kappa$,
however, complicate the interpretation of the apparent discrepancy
between the gas and dust radii \citep{2008Hughes}.
\citet{2019Trapman} suggest that a factor of 4 larger gas radii
is required to unambiguously prove the radial drift of dust.
On the other hand, simultaneous radiative transfer modeling of
gas and dust disks show that the same gas and dust radii with the
constant gas-to-dust ratio cannot reproduce the CO and dust intensity
profiles and the difference between them
\citep{2009Panic,Andrews2012,Ansdel2018,2019Facchini}.
These results imply presence of dust radial drift in the Class II stage.
In the case of the Class I protostar IRS 7B, we make use of the
simple observational results that the molecular line is severely
suppressed due to the strong dust emission in the dust disk region,
and that the gas emission is only visible at the outskirt of the dust disk.
To reproduce such observational images the gas disk radius larger than
the dust radius should be incorporated.
This implies that dust grains may have
migrated radially inward with respect to the molecular gas in the Class I disk,
even though they are not settled onto the disk midplane yet.


\textcolor{black}{The above conclusion may seem surprising at first sight. In the simplest case of a smooth laminar disk, the time scale for radial migration of the dust relative to the gas:
\begin{equation}
    t_{rad}\equiv \frac{R}{\vert v_{r,d}-v_{r,{\rm gas}}\vert} \approx \frac{1}{\eta\ \tau_s\ \Omega_K}
    \label{eq:RadialMigration}
\end{equation}
is much longer than the time scale for the dust to settle vertically:
\begin{equation}
    t_{z}\equiv \frac{z}{\vert v_{z,d}\vert} \approx \frac{1}{\tau_s\ \Omega_K}
    \label{eq:VerticalSettling}
\end{equation}
in the small particle limit \citep[with the dimensionless Stokes number $\tau_s\ll 1$, see Sections 7.1 and 7.2 of the review article by][]{2017atp..prop..107A}, where $R$ is the cylindrical radius, $v_{r,d}$ and $v_{r,{\rm gas}}$ the radial velocity of the dust and gas respectively, $\Omega_K$ the local Keplerian frequency, $z$ the vertical distance from the midplane, $v_{z,d}$ the dust settling terminal speed, and $\eta\propto (h/R)^2 \ll 1$ measures the fractional deviation of the gas rotation speed from the local Keplerian value. A potential solution to this apparent contradiction is that the dust is prevented from settling by turbulence in the disk, which may also be required to drive the relatively high mass accretion rate inferred from the dust continuum modeling. The nature of the turbulence and how it may affect the radial dust migration remain unclear and deserve further investigation.
In the circumternary dust ring around the GG Tau system,
\citet{2023Tang} proposed that the angular difference of the polarization
direction from the tangent of the ring could be due to the radial flow of
the dust grains. The radial drift velocity estimated from the angle
difference is, however, $\sim$3 times faster than the theoretical
prediction using the same formula by \citet{2017atp..prop..107A}.
}



\section{Summary}
\label{sec:sum}

We have constructed a physical model that reproduces the
1.3-mm dust-continuum emission and the C$^{18}$O (2--1) emission toward the Class I protostar R~CrA~IRS~7B-a, observed
with our ALMA Large Program, eDisk.
The gas and dust density
distributions are given, and thermal Monte Carlo simulations are
performed to calculate the spatial distribution of the dust temperature with RADMC-3D.
From the spatial distributions of gas and dust as well as the calculated temperature distribution, the images of the 1.3-mm dust-continuum emission and C$^{18}$O (2--1) emission are made with RADMC-3D,
followed by the observing simulations.
A number of such images were
created to search by eye for the model parameters that
decently reproduce the
intensities and the spatial
and velocity distributions of the observed emissions.
Our modeling effort has found the following insights into the physical
properties of the protostellar disk.

\begin{itemize}
    \item[1.] Dust disk models passively heated by the central protostar
    solely cannot
    reproduce the observed bright intensity of the 1.3-mm continuum emission, regardless of the adopted dust mass opacity table and the disk mass.
    The 1.3-mm dust-continuum emission in the disk is optically thick, and
    a change of $\kappa_{\rm 1.3mm}$ or disk mass does not increase the 1.3 mm intensity significantly. The 1.3-mm intensity in such models is
    a factor $\gtrsim$2 lower than the observed intensity.
    We found that the inclusion of the viscous accretion heating in the disk with a mass accretion rate of $\dot{M} \sim$10$^{-6}$ $M_{\odot}$ ${\rm yr}^{-1}$
    can reproduce the observed 1.3-mm intensities.
    Several other eDisk targets also exhibit similar high brightness temperatures of the 1.3-mm dust continuum
    emission originating from the disk. These results suggest that the protostellar disks are self-luminous.
    In contrast, disks around
    Class II sources do not show such high dust intensities,
    suggesting a difference in the physical conditions between the protostellar and Class II disks.

    \item[2.] The observed asymmetry of the 1.3-mm dust continuum image along the minor
    axis is reproduced with a flared dust disk model, where the flaring
    power-law index $q \sim$0.3 as defined by $\frac{h}{r} \sim r^{q}$
    and the disk mass $\sim$14 $\%$ of the central protostellar mass.
    The flaring index and disk mass are degenerate in reproducing
    the observed skewed intensity profile along the disk minor axis.
    This is because the $\tau$ = 1 disk surface is determined by
    the dust mass distribution, which in turn is set by the total dust mass and flaring parameter.
    Similar asymmetric structures of the dust emission along the disk minor axes
    are seen toward several other eDisk protostellar disks,
    suggesting that the dust in these protostellar disks is yet to settle to
    the midplane. This is different from the geometrically-thin dust distribution
    in the disks around Class II sources associated with ring/gap features.
    These results suggest that dust settling and formation of ring/gap features
    in the disks proceed between the Class I and Class II stages.

    \item[3.] To reproduce the observed distribution of the
    C$^{18}$O emission, the radius of the gas disk should be
    $\sim$80~au, i.e., larger than that of the dust disk radius
    ($\sim$60~au). In the models with the gas disk radius identical
    to the dust disk radius, the molecular emission is severely suppressed
    because of the obscuration by the high brightness and opacity of the dust emission.
    
    \item[4.] Our radiative transfer model, 
    combined with the observing simulations, reproduces the observed depression of the C$^{18}$O emission toward the
    continuum emission. Around the systemic velocity, the
    observed absorption of the C$^{18}$O line is also
    reproduced with our model. The absorption feature should be
    originated from the higher background dust brightness temperature
    than the line excitation temperature. The depression of
    the molecular emission is also caused by
    the higher dust optical depth than the line optical depth.
    Note that gas components are present throughout the disk, with a constant C$^{18}$O abundance,
    in our models. While the apparent depressions of the molecular emission in the disks
    observed with ALMA are often interpreted as a real depletion of the molecule, they could be caused by radiative transfer effects, which are further enhanced by interferometric filtering. Detailed radiative transfer modeling and interferometric observing simulations are required to interpret interferometric images correctly.
\end{itemize}

We would like to thank all the
ALMA staff supporting this work. S.T. is supported by JSPS
KAKENHI grant Nos. JP21H00048 and JP21H04495 and by NAOJ
ALMA Scientific Research grant No. 2022-20A. K.S. is
supported by JSPS KAKENHI grant No. JP21H04495. N.O.
acknowledges support from National Science and Technology
Council (NSTC) in Taiwan through grants
NSTC 109-2112-M-001-051 and 110-2112-M-001-031.
J.J.T. acknowledges support
from NASA RP 80NSSC22K1159. J.K.J., R.S., and S.G. acknowledge support
from the Independent Research Fund Denmark
(grant No. 0135-00123B).
Y.A. acknowledges support by NAOJ ALMA Scientific
Research Grant code 2019-13B, Grant-in-Aid for Scientific
Research (S) JP18H05222, and Grant-in-Aid for Transformative
Research Areas (A) JP20H05844 and JP20H05847. F.J.E. acknowledges
support from NSF AST-2108794. I.D.G.-M. acknowledges support from grant
PID2020-114461GB-I00,
funded by MCIN/AEI/10.13039/501100011033.
P.M.K. acknowledges support from NSTC 108-2112-M-001-012,
NSTC 109-2112-M-001-022, and NSTC 110-2112-M-001-057. W.K. was supported by the National Research
Foundation of Korea (NRF) grant funded by the Korea
government (MSIT) (NRF-2021R1F1A1061794). S.-P.L. and T.
J.T. acknowledge grants from the NSTC of Taiwan
106-2119-M-007-021-MY3 and 109-2112-M-007-010-MY3.
C.W.L. is supported
by the Basic Science Research Program through the NRF
funded by the Ministry of Education, Science and Technology
(NRF-2019R1A2C1010851) and by the Korea Astronomy and
Space Science Institute grant funded by the Korea government
(MSIT; project No. 2022-1-840-05). J.-E.L. is supported by the
NRF grant funded by the Korean government (MSIT; grant No.
2021R1A2C1011718). Z.-Y.L. is supported in part by NASA
NSSC20K0533 and NSF AST-2307199 and AST-1910106. Z.-Y.D.L. acknowledges
support from the Jefferson Scholars Foundation, the NRAO
ALMA Student Observing Support (SOS) SOSPA8-003, the
Achievements Rewards for College Scientists (ARCS) Foundation
Washington Chapter, the Virginia Space Grant Consortium
(VSGC), and UVA research computing (RIVANNA). L.W.L. and F.J.E. acknowledge support from NSF AST-2108794. S.M. is
supported by JSPS KAKENHI grant Nos. JP21J00086 and
JP22K14081. S.N. acknowledges support from the National Science
Foundation through the Graduate Research Fellowship Program
under grant No. 2236415. 
P.D.S. acknowledges support from NSF AST-2001830 and NSF
AST-2107784.
K.T. is supported by JSPS KAKENHI grant Nos. JP21H04487, JP22KK0043,
and JP21H04495.
M.L.R.H. acknowledges support from the Michigan Society of Fellows. J.P.W. acknowledges support from NSF
AST-2107841. Y.Y. is supported by the International Graduate
Program for Excellence in Earth-Space Science (IGPEES),
World-leading Innovative Graduate Study (WINGS) Program of
the University of Tokyo. H.-W.Y. acknowledges support from the
NSTC in Taiwan through grant NSTC 110-2628-M-001-003-MY3 and from the Academia Sinica Career Development Award
(AS-CDA-111-M03). This paper makes use of the following
ALMA data: ADS/JAO.ALMA $\#$2019.1.00261.L and
2019.A.00034.S. ALMA is a partnership of ESO (representing its
member states), NSF (USA), and NINS (Japan), together with
NRC (Canada), NSTC and ASIAA (Taiwan), and KASI
(Republic of Korea), in cooperation with the Republic of Chile.
The Joint ALMA Observatory is operated by ESO, AUI/NRAO,
and NAOJ. The National Radio Astronomy Observatory is a
facility of the National Science Foundation operated under
cooperative agreement by Associated Universities, Inc.

\appendix
\section{Searched Model Prameters}

Table \ref{search} lists all the sets of the calculated
model parameters, sorted by the disk flaring index $q$ and then the disk mass $M_d$,
except for models 100-102 which adopt $simobserve$ instead of $ft$.
Note that we did not attempt to cover the complete parameter space, but tried to approach
to the model images which reproduce the observed images.

\begin{longrotatetable}
\begin{deluxetable}{cccccccccc}
\tablecaption{Parameter Search for the IRS 7B-a modeling \label{search}}
\tabletypesize{\scriptsize}
\tablewidth{250pt} 
\tablehead{\colhead{Model} & \colhead{Grid}&\colhead{Flaring Index, $q$} & \colhead{$M_{d}$} & \colhead{$r_{dust}$} & \colhead{$r_{gas}$} &  \colhead{Mass Accretion Rate $\dot{M}$} &
\colhead{Dust Opacity} & \colhead{Disk inclination} & \colhead{Stellar Luminosity} \\
\colhead{} &\colhead{($r,\theta,\phi$)} & \colhead{} & \colhead{($M_\odot$)} &
\colhead{(au)} &\colhead{(au)} & \colhead{($\times~10^{-6}~M_{\odot}~yr^{-1}$)} & \colhead{}  &\colhead{($\degr$)}  & \colhead{($L_\odot$)}}
\startdata
1	&	(512, 512, 1)	&	0.00	&	0.39	&	60	&	80	&	0.0	&	Beckwith 	&	-70	&	5.2	\\
2	&	(512, 512, 1)	&	0.00	&	0.39	&	60	&	80	&	1.9	&	Beckwith 	&	-70	&	5.2	\\
3	&	(512, 512, 1)	&	0.00	&	0.39	&	60	&	80	&	1.4	&	Beckwith 	&	-70	&	5.2	\\
4	&	(512, 512, 1)	&	0.00	&	0.41	&	62	&	80	&	1.4	&	DSHARP	&	-70	&	5.2	\\
5	&	(512, 512, 1)	&	0.00	&	0.55	&	62	&	80	&	1.4	&	DSHARP	&	-70	&	5.2	\\
6	&	(512, 512, 1)	&	0.20	&	0.39	&	60	&	80	&	1.4	&	Beckwith 	&	-70	&	5.2	\\
7	&	(256, 256, 1)	&	0.20	&	0.39	&	60	&	80	&	1.9	&	Beckwith 	&	-70	&	5.2	\\
8	&	(512, 512, 1)	&	0.20	&	0.41	&	62	&	80	&	1.4	&	DSHARP	&	-70	&	5.2	\\
9	&	(256, 256, 1)	&	0.20	&	0.45	&	60	&	80	&	1.9	&	Beckwith 	&	-70	&	5.2	\\
10	&	(512, 512, 1)	&	0.20	&	0.55	&	62	&	80	&	1.4	&	DSHARP	&	-70	&	5.2	\\
11	&	(512, 512, 1)	&	0.20	&	0.68	&	62	&	80	&	1.4	&	DSHARP	&	-70	&	5.2	\\
12	&	(256, 256, 1)	&	0.25	&	0.03	&	60	&	80	&	5.1	&	Beckwith 	&	-70	&	5.2	\\
13	&	(256, 256, 1)	&	0.25	&	0.06	&	60	&	80	&	2.9	&	Beckwith 	&	-70	&	5.2	\\
14	&	(256, 256, 1)	&	0.25	&	0.13	&	60	&	80	&	2.3	&	Beckwith 	&	-70	&	5.2	\\
15	&	(256, 256, 1)	&	0.25	&	0.19	&	60	&	80	&	2.0	&	Beckwith 	&	-70	&	5.2	\\
16	&	(512, 512, 1)	&	0.25	&	0.20	&	62	&	80	&	1.4	&	DSHARP	&	-70	&	5.2	\\
17	&	(256, 256, 1)	&	0.25	&	0.26	&	60	&	80	&	1.9	&	Beckwith 	&	-70	&	5.2	\\
18	&	(512, 512, 1)	&	0.25	&	0.27	&	62	&	80	&	1.4	&	DSHARP	&	-70	&	5.2	\\
19	&	(256, 256, 1)	&	0.25	&	0.32	&	60	&	80	&	1.9	&	Beckwith 	&	-70	&	5.2	\\
20	&	(512, 512, 1)	&	0.25	&	0.39	&	60	&	80	&	1.4	&	Beckwith 	&	-70	&	5.2 \\
21	&	(512, 512, 1)	&	0.25	&	0.39	&	60	&	80	&	1.9	&	Beckwith 	&	-70	&	5.2	\\
22	&	(512, 512, 1)	&	0.25	&	0.39	&	60	&	80	&	1.9	&	DSHARP	&	-70	&	5.2	\\
23	&	(512, 512, 1)	&	0.25	&	0.39	&	60	&	80	&	1.9	&	Beckwith 	&	-70	&	20.0	\\
24	&	(512, 512, 1)	&	0.25	&	0.39	&	60	&	80	&	1.9	&	Beckwith 	&	-72	&	20.0	\\
25	&	(512, 512, 1)	&	0.25	&	0.39	&	60	&	80	&	1.9	&	Beckwith 	&	-74	&	20.0	\\
26	&	(512, 512, 1)	&	0.25	&	0.39	&	60	&	80	&	0.0	&	Beckwith 	&	-70	&	5.2	\\
27	&	(512, 512, 1)	&	0.25	&	0.39	&	60	&	80	&	0.0	&	Beckwith 	&	-70	&	20.0	\\
28	&	(512, 512, 1)	&	0.25	&	0.41	&	62	&	80	&	1.4	&	DSHARP	&	-70	&	5.2	\\
29	&	(512, 512, 1)	&	0.25	&	0.45	&	60	&	80	&	1.7	&	Beckwith 	&	-70	&	5.2	\\
30	&	(512, 512, 1)	&	0.25	&	0.45	&	60	&	80	&	1.7	&	Beckwith 	&	-72	&	5.2	\\
31	&	(512, 512, 1)	&	0.25	&	0.45	&	60	&	80	&	1.7	&	Beckwith 	&	-75	&	5.2	\\
32	&	(256, 256, 1)	&	0.25	&	0.45	&	60	&	80	&	1.9	&	Beckwith 	&	-70	&	5.2	\\
33	&	(512, 512, 1)	&	0.25	&	0.55	&	62	&	80	&	1.4	&	DSHARP	&	-70	&	5.2	\\
34	&	(512, 512, 1)	&	0.25	&	0.55	&	62	&	80	&	0.0	&	DSHARP	&	-70	&	5.2	\\
35	&	(512, 512, 1)	&	0.25	&	0.68	&	62	&	80	&	1.4	&	DSHARP	&	-70	&	5.2	\\
36	&	(256, 256, 1)	&	0.30	&	0.06	&	60	&	80	&	2.9	&	Beckwith 	&	-70	&	5.2	\\
37	&	(256, 256, 1)	&	0.30	&	0.13	&	60	&	80	&	2.3	&	Beckwith 	&	-70	&	5.2	\\
38	&	(256, 256, 1)	&	0.30	&	0.19	&	60	&	80	&	2.0	&	Beckwith 	&	-70	&	5.2	\\
39	&	(512, 512, 1)	&	0.30	&	0.20	&	62	&	80	&	1.4	&	DSHARP	&	-70	&	5.2	\\
40	&	(256, 256, 1)	&	0.30	&	0.26	&	60	&	80	&	2.0	&	Beckwith 	&	-70	&	5.2	\\
41	&	(512, 512, 1)	&	0.30	&	0.27	&	62	&	80	&	1.4	&	DSHARP	&	-70	&	5.2	\\
42	&	(512, 512, 1)	&	0.30	&	0.32	&	60	&	80	&	1.9	&	Beckwith 	&	-68	&	5.2	\\
43	&	(512, 512, 1)	&	0.30	&	0.32	&	60	&	80	&	1.9	&	Beckwith 	&	-70	&	5.2	\\
44	&	(512, 512, 1)	&	0.30	&	0.32	&	60	&	80	&	1.9	&	Beckwith 	&	-72	&	5.2	\\
45	&	(512, 512, 1)	&	0.30	&	0.39	&	60	&	80	&	1.4	&	Beckwith 	&	-70	&	5.2	\\
46	&	(512, 512, 1)	&	0.30	&	0.39	&	60	&	80	&	1.3	&	Beckwith 	&	-70	&	5.2	\\
47	&	(512, 512, 1)	&	0.30	&	0.39	&	60	&	80	&	1.2 &	Beckwith 	&	-70	&	5.2	\\
48	&	(512, 512, 1)	&	0.30	&	0.39	&	60	&	80	&	1.2	&	DSHARP	&	-70	&	5.2	\\
49	&	(512, 512, 1)	&	0.30	&	0.39	&	60	&	80	&	0.0	&	Beckwith 	&	-70	&	26.0	\\
50	&	(512, 512, 1)	&	0.30	&	0.39	&	60	&	80	&	0.0	&	Beckwith 	&	-70	&	5.2	\\
51	&	(512, 512, 1)	&	0.30	&	0.39	&	60	&	80	&	1.5	&	Beckwith 	&	-70	&	5.2	\\
52	&	(512, 512, 1)	&	0.30	&	0.39	&	60	&	80	&	1.6	&	Beckwith 	&	-70	&	5.2	\\
53	&	(512, 512, 1)	&	0.30	&	0.39	&	60	&	80	&	1.7	&	Beckwith 	&	-70	&	5.2	\\
54	&	(512, 512, 1)	&	0.30	&	0.39	&	60	&	80	&	1.8	&	Beckwith 	&	-70	&	5.2	\\
55	&	(512, 512, 1)	&	0.30	&	0.39	&	60	&	80	&	1.7	&	Beckwith 	&	-70	&	5.2	\\
56	&	(512, 512, 1)	&	0.30	&	0.39	&	60	&	80	&	1.9	&	Beckwith 	&	-70	&	5.2	\\
57	&	(512, 512, 1)	&	0.30	&	0.39	&	60	&	80	&	1.9	&	DSHARP	&	-70	&	5.2	\\
58	&	(512, 512, 1)	&	0.30	&	0.39	&	60	&	80	&	1.9	&	DSHARP	&	-72	&	5.2	\\
59	&	(512, 512, 1)	&	0.30	&	0.39	&	60	&	80	&	0.0	&	Beckwith 	&	-70	&	5.2	\\
60	&	(512, 512, 1)	&	0.30	&	0.39	&	60	&	80	&	0.0	&	Beckwith 	&	-70	&	52.0	\\
61 (Fiducial)	&	(512, 512, 1)	&	0.30	&	0.41	&	62	&	80	&	1.4	&	DSHARP	&	-70	&	5.2	\\
62	&	(512, 512, 1)	&	0.30	&	0.41	&	62	&	80	&	1.4	&	DSHARP	&	-72	&	5.2	\\
63	&	(512, 512, 1)	&	0.30	&	0.41	&	62	&	80	&	1.4	&	Beckwith 	&	-70	&	5.2	\\
64	&	(512, 512, 1)	&	0.30	&	0.41	&	62	&	80	&	1.4	&	OH5	&	-70	&	5.2	\\
65	&	(512, 512, 1)	&	0.30	&	0.41	&	62	&	80	&	1.4	&	Semenov03	&	-70	&	5.2	\\
66	&	(512, 512, 1)	&	0.30	&	0.41	&	62	&	80	&	0.0	&	DSHARP	&	-70	&	5.2	\\
67	&	(512, 512, 1)	&	0.30	&	0.41	&	62	&	80	&	0.0	&	DSHARP	&	-70	&	26.0	\\
68	&	(512, 512, 1)	&	0.30	&	0.41	&	62	&	80	&	0.0	&	Beckwith 	&	-70	&	5.2	\\
69	&	(512, 512, 1)	&	0.30	&	0.41	&	62	&	80	&	0.0	&	Semenov03	&	-70	&	5.2	\\
70	&	(512, 512, 1)	&	0.30	&	0.41	&	62	&	80	&	0.0	&	OH5	&	-70	&	5.2	\\
71	&	(512, 512, 1)	&	0.30	&	0.45	&	60	&	80	&	1.9	&	Beckwith 	&	-70	&	5.2	\\
72	&	(512, 512, 1)	&	0.30	&	0.45	&	60	&	80	&	1.9	&	DSHARP	&	-70	&	5.2	\\
73	&	(512, 512, 1)	&	0.30	&	0.55	&	62	&	80	&	1.4	&	DSHARP	&	-70	&	5.2	\\
74	&	(512, 512, 1)	&	0.30	&	0.60	&	62	&	62	&	1.4	&	DSHARP	&	-70	&	5.2	\\
75	&	(512, 512, 1)	&	0.30	&	0.60	&	60	&	60	&	1.4	&	Beckwith 	&	-70	&	5.2	\\
76	&	(512, 512, 1)	&	0.30	&	0.60	&	80	&	80	&	1.4	&	DSHARP	&	-70	&	5.2	\\
77	&	(512, 512, 1)	&	0.30	&	0.60	&	60	&	60	&	1.9	&	Beckwith 	&	-70	&	5.2	\\
78	&	(512, 512, 1)	&	0.30	&	0.68	&	62	&	80	&	1.4	&	DSHARP	&	-70	&	5.2	\\
79	&	(512, 512, 1)	&	0.30	&	0.80	&	60	&	60	&	1.4	&	OH5	&	-70	&	5.2	\\
80	&	(512, 512, 1)	&	0.30	&	0.80	&	60	&	60	&	1.4	&	Semenov03	&	-70	&	5.2	\\
81	&	(256, 256, 1)	&	0.35	&	0.06	&	60	&	80	&	4.0	&	Beckwith 	&	-70	&	5.2	\\
82	&	(512, 512, 1)	&	0.35	&	0.27	&	62	&	80	&	1.4	&	DSHARP	&	-70	&	5.2	\\
83	&	(512, 512, 1)	&	0.35	&	0.32	&	60	&	80	&	1.9	&	Beckwith 	&	-70	&	5.2	\\
84	&	(512, 512, 1)	&	0.35	&	0.39	&	60	&	80	&	1.4	&	Beckwith 	&	-70	&	5.2	\\
85	&	(512, 512, 1)	&	0.35	&	0.41	&	62	&	80	&	1.4	&	DSHARP	&	-70	&	5.2	\\
86	&	(512, 512, 1)	&	0.35	&	0.55	&	62	&	80	&	1.4	&	DSHARP	&	-70	&	5.2	\\
87	&	(256, 256, 1)	&	0.40	&	0.03	&	60	&	80	&	5.0	&	Beckwith 	&	-70	&	5.2	\\
88	&	(256, 256, 1)	&	0.40	&	0.06	&	60	&	80	&	2.9	&	Beckwith 	&	-70	&	5.2	\\
89	&	(256, 256, 1)	&	0.40	&	0.13	&	60	&	80	&	2.3	&	Beckwith 	&	-70	&	5.2	\\
90	&	(256, 256, 1)	&	0.40	&	0.19	&	60	&	80	&	2.1	&	Beckwith 	&	-70	&	5.2	\\
91	&	(256, 256, 1)	&	0.40	&	0.19	&	60	&	80	&	0.0	&	Beckwith 	&	-70	&	5.2	\\
92	&	(256, 256, 1)	&	0.40	&	0.26	&	60	&	80	&	2.2	&	Beckwith 	&	-70	&	5.2	\\
93	&	(512, 512, 1)	&	0.40	&	0.27	&	62	&	80	&	1.4	&	DSHARP	&	-70	&	5.2	\\
94	&	(512, 512, 1)	&	0.40	&	0.41	&	62	&	80	&	1.4	&	DSHARP	&	-70	&	5.2	\\
95	&	(512, 512, 1)	&	0.45	&	0.27	&	62	&	80	&	1.4	&	DSHARP	&	-70	&	5.2	\\
96	&	(256, 256, 1)	&	0.50	&	0.06	&	60	&	80	&	3.0	&	Beckwith 	&	-70	&	5.2	\\
97	&	(256, 256, 1)	&	0.50	&	0.13	&	60	&	80	&	2.4	&	Beckwith 	&	-70	&	5.2	\\
98	&	(256, 256, 1)	&	0.50	&	0.19	&	60	&	80	&	2.0	&	Beckwith 	&	-70	&	5.2 \\
99	&	(256, 256, 1)	&	0.60	&	0.06	&	60	&	80	&	3.0	&	Beckwith 	&	-70	&	5.2 \\
100 (simobserve) &   (512, 512, 1)   &	0.30	&	0.41	&	62	&	80	&	1.4	&	DSHARP      &	-70	&	5.2	\\
101 (simobserve) &   (512, 512, 1)   &	0.30	&	0.41	&	62	&	62	&	1.4	&	DSHARP      &	-70	&	5.2	\\
102 (simobserve) &   (512, 512, 1)   &	0.30	&	0.41	&	80	&	80	&	1.4	&	DSHARP      &	-70	&	5.2	\\
\enddata
\end{deluxetable}
\end{longrotatetable}


\software{CASA \citep{Mcmullin2007}, matplotlib \citep{Hunter2007}, RADMC-3D \citep{2012Dullemond}, bettermoments \citep{2018Teague,2019Teague}, PVextractor \citep{2016Ginsburg}, APLpy \citep{aplpy2012,aplpy2019}, SLAM \citep{yusuke_aso_2023_7783868}, astropy \citep{2022Astropy}}
\facility{ALMA}

\bibliography{takakuwa}{}
\bibliographystyle{aasjournal}



\end{document}